\def\Re{{\rm Re}}
\def\Im{{\rm Im}}
\def\PiL{\Pi_{\rm L}}
\def\PiT{\Pi_{\rm T}}

\def\PT{P_{\rm T}}
\def\ret{{\rm Ret}}
\def\adv{{\rm Adv}}
\def\sgn{{\rm sign}}

\def\ppar{{p_\parallel}}
\def\qpar{{q_\parallel}}
\def\ipa{inelastic pair annihilation}
\def\LPM{{\scriptscriptstyle \rm LPM}}

\def\x{{\bf x}}
\def\p{{\bf p}}
\def\q{{\bf q}}
\def\k{{\bf k}}

\def\cf{C_{\rm F}}

\def\df{d_{\rm F}}

\def\nc{N_{\rm c}}
\def\mD{m_{\rm D}}

\def\Eqs{Eqs.~}
\def\barn{\overline n}
\def\bG{\overline G}

\newcommand\Eq[1]{Eq.~(\ref{#1})}
\def\hsp  {\hspace{ 0.27em}}
\def\nhsp {\hspace{-0.27em}}
\def\hspp {\hspace{ 0.18em}}
\def\nhspp{\hspace{-0.18em}}
\newcommand{\nott}[1]{{\hspp\not{\nhsp #1\hsp}\nhspp}}

\def\half{{\textstyle{1\over2}}}
\def\gs{g_{\rm s}}

\def\Ns{N_{\rm s}}
\def\Nf{N_{\rm f}}

\def\alphas{\alpha_{\rm s}}
\def\alphaEM{\alpha_{\scriptscriptstyle \rm EM}}
\def\bSigma{\mbox{\boldmath$\Sigma$}}

\def\gsim{\mbox{~{\raisebox{0.4ex}{$>$}}\hspace{-1.1em}
	{\raisebox{-0.6ex}{$\sim$}}~}}
\def\lsim{\mbox{~{\raisebox{0.4ex}{$<$}}\hspace{-1.1em}
	{\raisebox{-0.6ex}{$\sim$}}~}}

\newfam\scrfam                                          
\global\font\twelvescr=rsfs10 scaled\magstep1%
\global\font\eightscr=rsfs7 scaled\magstep1%
\global\font\sixscr=rsfs5 scaled\magstep1%
\skewchar\twelvescr='177\skewchar\eightscr='177\skewchar\sixscr='177%
\textfont\scrfam=\twelvescr\scriptfont\scrfam=\eightscr
\scriptscriptfont\scrfam=\sixscr

\documentstyle[preprint,aps,fixes,epsf,eqsecnum]{revtex}

\makeatletter           
\@floatstrue
\def\figure{\let\@capwidth\columnwidth\@float{figure}}
\let\endfigure\end@float
\@namedef{figure*}{\let\@capwidth\textwidth\@dblfloat{figure}}
\@namedef{endfigure*}{\end@dblfloat}
\def\table{\let\@capwidth\columnwidth\@float{table}}
\let\endtable\end@float
\@namedef{table*}{\let\@capwidth\textwidth\@dblfloat{table}}
\@namedef{endtable*}{\end@dblfloat}
\def\la{\label}
\makeatother

\tightenlines
\newcommand\pcite[1]{\protect{\cite{#1}}}

\begin {document}


\preprint {UW/PT 01--21}

\title
    {
    Photon Emission from Ultrarelativistic Plasmas
    }

\author {Peter Arnold}
\address
    {%
    Department of Physics,
    University of Virginia,
    Charlottesville, VA 22901
    }%
\author{Guy D. Moore and Laurence G. Yaffe}
\address
    {%
    Department of Physics,
    University of Washington,
    Seattle, WA 98195
    }%

\date {September 2001} 

\maketitle
\vskip -20pt

\begin {abstract}%
    {%
    The emission rate of photons from a hot,
    weakly coupled ultrarelativistic plasma is analyzed.
    Leading-log results, reflecting the sensitivity of the emission rate
    to scattering events with momentum transfers from $gT$ to $T$,
    have previously been obtained.
    But a complete leading-order treatment requires including
    collinearly enhanced, inelastic processes such as bremsstrahlung.
    These inelastic processes receive 
    $O(1)$ modifications from multiple scattering during the photon emission
    process, which limits the coherence length of the emitted radiation 
    (the Landau-Pomeranchuk-Migdal effect).
    We perform a diagrammatic analysis to identify,
    and sum, all leading-order contributions.
    We find that the leading-order photon emission rate is
    not sensitive to non-perturbative $g^2 T$ scale dynamics.
    We derive an integral equation for the photon emission rate
    which is very similar to the result of Migdal in his
    original discussion of the LPM effect.
    The accurate solution of this integral equation for
    specific theories of interest will be reported in a
    companion paper.
    }%
\end {abstract}

\thispagestyle{empty}

\section{Introduction and Results}
\la{sec:intro}

The recent commissioning of RHIC,
and anticipated heavy ion experiments at the LHC,
have stimulated work on photon emission in both equilibrium and
nonequilibrium relativistic plasmas
\cite {softgamma,Gelis1,Gelis2,Gelis3,o1,o2,o3,o4,o5,Baier_QED,Zakharov_QED}.
At sufficiently high temperatures or energy densities,
QCD interactions become weak,
and perturbation theory can be applicable
for observables which are predominantly sensitive to the dynamics
of typical excitations in the plasma, those with momenta
comparable to the temperature.
Whether perturbation theory is really useful for the energy
densities achievable at RHIC and the LHC is an open question,
but thermal field theory also has important applications in
early universe cosmology (such as baryogenesis, leptogenesis,
and magnetogenesis).
From a purely theoretical perspective, hot, weakly coupled,
relativistic plasmas provide an instructive and interesting
domain with rich dynamics dependent on the interplay of a
wide variety of quantum and statistical effects,
in which one can, with effort, perform controlled calculations
for physically interesting quantities.

\subsection {Background}

The basic perturbative tools for studying a relativistic plasma near
equilibrium were developed long ago \cite{Keldysh}.  In the last
decade or so, there has been substantial progress.
It is well understood today that there are three distinct
length scales on which a hot weakly coupled relativistic plasma
exhibits different characteristic behavior.  

The first scale is the ``hard'' scale,
corresponding to wavenumbers (or momenta) $k \sim T$.
This is the characteristic scale of momentum or energy for
the vast majority of the excitations (quarks, gluons, ...)
comprising the plasma.
The contributions of such excitations dominate bulk thermodynamic properties
such as the total energy and momentum density of the plasma,
as well as conserved charge susceptibilities
(electric charge and various conserved or approximately
conserved flavor charges, such as baryon number and isospin).
The behavior of these excitations also controls
transport properties such as viscosity, baryon number diffusion,
and electrical conductivity \cite {transport2,HosoyaKajantie,BMPRa}.
Provided the gauge coupling%
\footnote
    {%
    Here, and throughout our discussion,
    the gauge coupling is to be understood as
    the relevant coupling defined at the scale of the temperature,
    $g \equiv g(T)$.
    For hot QCD, $g$ means the strong gauge coupling $\gs$,
    while for hot QED, $g$ should be understood as the electric charge $e$.
    Our treatment is applicable to both cases,
    although a few of the following comments concerning $g^2 T$ scale
    dynamics assume that the gauge field is non-Abelian.
    }
is weak,
$g \ll 1$,
excitations with wave number $k \sim T$
propagate as almost free quasi-particles moving at essentially
the speed of light, engaging in
occasional small angle scatterings
[deflections by an angle of $O(g)$ or less]
with a mean free time of order $1/g^2 T \gg 1/T$,
and even more occasional large angle scatterings [with $O(1)$ deflection]
on times scales of order $1/g^4 T$.
These parametric estimates ignore factors of $\ln g^{-1}$
which are present in both mean free times
\cite {HosoyaKajantie,BMPRa,damping-rates,jj}.

The second scale is the ``soft'' (or ``semi-hard'') scale, $k \sim gT$.
For excitations of this wave number, plasma effects such as Debye screening,
plasma oscillations, and Landau damping, become of $O(1)$ importance.
These arise because of coherent interactions of the $gT$ scale fields
with the thermal bath of harder, $O(T)$ scale excitations. 
Dispersion relations, and non-Abelian gauge interactions,
receive $O(1)$ corrections which must be appropriately resummed
into the propagators and vertices
\cite{HTL1,HTL2,HTL3}.
This resummation is commonly referred to as ``hard-thermal-loop'' (HTL)
perturbation theory.
After this selective resummation,
the effective interactions amongst the $gT$ scale degrees of freedom are weak,
and hence perturbation theory remains applicable for $gT$ scale dynamics.
However, perturbation theory at the $gT$ scale is only an expansion in $g$,
not $g^2$.
The primary physical importance of $gT$ excitations comes from the role
they play as exchanged particles in scattering processes.
The ``transport'' (or large-angle) scattering rate is dominated by
exchange momenta in the range from $T$ down to $gT$,
while the total scattering rate
is dominated by exchange momenta in the range from $gT$ down to $g^2 T$.

The third scale is the ``ultrasoft'' or non-perturbative scale, 
$k \sim g^2 T$.
For gauge bosons (or scalars, if the temperature is
sufficiently close to a phase transition temperature), 
the occupation number of individual modes at this scale
is so large, $O(1/g^2)$,
that the mutual interactions between $g^2 T$ scale degrees of
freedom are truly strongly coupled; perturbative treatments break down.
This scale is important for problems which require non-perturbative
physics, such as baryon number violation in the hot standard model above
the electroweak phase transition \cite{sphaleron1}.

One would expect that questions involving the long time behavior of
sensible observables, dominated by $k \sim T$ excitations,
should only depend on $T$ and $gT$ scale physics,
at least at leading order in powers of $g$,
and should therefore be perturbatively computable.%
\footnote
    {%
    The total scattering rate for a hard excitation,
    with its logarithmic sensitivity to $g^2 T$ scale dynamics,
    is not really a meaningful observable,
    for roughly the same reason that the strictly elastic
    scattering amplitude for electrons in QED is not meaningful.
    }
Numerous physically interesting questions fall into this category.
Some examples are the hard photon emission rate from a thermalized
quark-gluon plasma,
the energy loss of a very hard parton (thermal ``jet quenching''), 
and transport coefficients such as shear and bulk viscosity,
electrical conductivity,
and various diffusion constants (baryon number, isospin, etc.).
In general, the question of how fast a nonequilibrium plasma relaxes
to equilibrium, and which degrees of freedom have the slowest approach to
equilibrium, depends on the dynamics of typical hard excitations.

\begin{figure}[t]
\centerline{\epsfxsize=5.5in\epsfbox{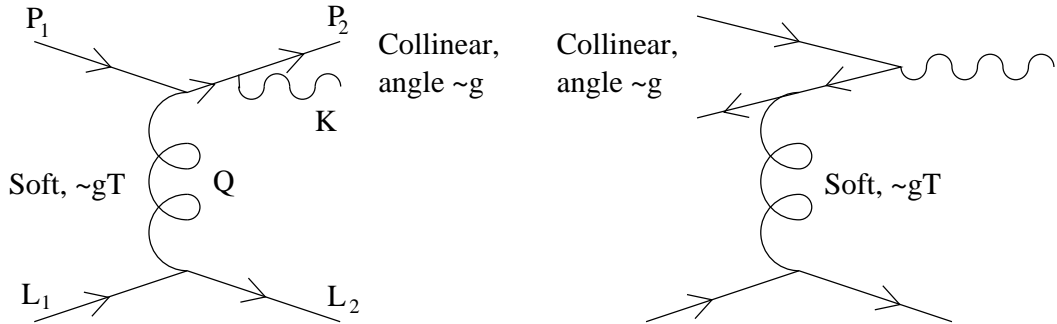}}
\vspace{0.1in}
\caption{\label{intro_fig}
The two processes of interest, bremsstrahlung and \ipa.
In each case, the emerging photon is hard, with energy $\sim T$,
but is nearly collinear with the quark from which it is radiated.
One or more interactions which exchange momentum with other excitations
in the plasma are required for these processes to occur;
the exchange momenta are all soft, with energy and momentum $\sim gT$.
Time should be viewed as running from left to right.
}
\end{figure}

Surprisingly, not one of the physical observables just mentioned
has been correctly computed, in a hot gauge theory, beyond leading order
in the {\em logarithm} of the coupling, $\log(1/g)$
(that is, neglecting corrections suppressed only by $1/\ln g^{-1}$).
The chief obstruction to such an evaluation is the fact that each of
these problems depends, at leading order in $g$, on the processes depicted in
Fig.~\ref{intro_fig}, namely bremsstrahlung and \ipa, with
soft exchange and nearly collinear external states.
These processes contribute to transport mean free paths and to
the photo-emission rate at leading order, despite
being $2 \leftrightarrow 3$ particle processes,
because a combination of soft and collinear enhancements make them
occur at an $O(g^4T^4)$ rate per spacetime volume
[or, for photo-emission from a quark gluon plasma, $O(e^2 \gs^2 T^4)$],
which is the same as $2 \leftrightarrow 2$ scattering processes.
The importance of these processes have been emphasized recently by
Aurenche {\em et~al.} \cite{Gelis1}, who demonstrated that they
contribute at leading order to photon emission
from the quark-gluon plasma.  
This was an important realization, but the quantitative analysis was
incomplete (as recognized by the same authors in \cite{Gelis2,Gelis3}),
because it ignored a suppression of the emission rate due to the
Landau-Pomeranchuk-Migdal (LPM) effect.
That is, one internal propagator in
the diagram is very nearly on-shell and receives $O(1)$ corrections
from the quasiparticle width.
This reflects sensitivity to additional scatterings which occur
{\em during} the photon emission process.
However, merely including the appropriate width on the intermediate
propagator is not sufficient; a consistent treatment requires a
more elaborate and detailed analysis.
The purpose of this paper is to provide this analysis,
and to derive an integral equation, similar in form to a linearized
kinetic equation, whose solution determines the rate of
bremsstrahlung and \ipa\ processes,
to leading order in $g$.
The accurate solution of this integral equation for
specific theories of interest will be reported in a
companion paper \cite {AMY2}.

At times in our discussion we will refer to the relevant charge
carriers as ``quarks'' and the exchanged gauge bosons as ``gluons'',
but it should be emphasized that our results apply equally
well to hot QED.
We assume that the relevant interactions in the plasma 
come from a weakly coupled simple gauge group,
either Abelian or non-Abelian.
We also assume that the emitted photon has negligible
interactions with the plasma subsequent to its emission.%
\footnote
    {%
    Of course, the propagation of a photon really is
    affected by the medium it traverses;
    in particular, thermal corrections to the photon dispersion
    relation increase the energy of a
    photon of momentum $\sim T$ by an amount of order $e^2 T$.
    In a hot QCD plasma, the photon coupling $e$ is
    small compared to the strong coupling constant $\gs$, so thermal
    corrections to the photon dispersion relation are negligible
    compared to the (non-electromagnetic) thermal effects we will
    be focusing on.
    For QED, instead of speaking of the photon emission rate,
    which ignores rescattering, one should consider the photon emissivity.
    Our analysis is also applicable in this case, although one would
    need to solve the final integral equation using a timelike photon
    4-momentum suitably determined by its thermal dispersion relation.
    }
Although it is also physically quite interesting,
we will not consider the case where the hard collinear emitted
particle is a gluon instead of a photon, but leave this for future work.

\subsection {Soft exchange with collinear emission}

Let us first indicate why the processes shown in Fig.~\ref{intro_fig}
are important, by sketching the evaluation of the bremsstrahlung rate;
estimating the pair annihilation rate is completely analogous.
Our argument roughly follows that of Aurenche {\em et~al.}~\cite{Gelis1}.
Assume, for simplicity, that the photon momentum is hard, $k \sim T$.
The analysis is easiest for scalar quarks,
although the conclusions we will find also hold for real fermionic quarks.
Labeling the momenta as shown in the figure,%
\footnote
    {%
    As is customary in thermal field theory, we work in the plasma rest
    frame, write 4-momenta in capitals ($Q$), their spatial components
    in bold face $(\q)$, the magnitude of the three-momentum in lower
    case ($q\equiv |\q|$), and the time component in lower case 
    with zero index ($q^0$).
    }
and requiring that the outgoing states be on-shell fixes $K^2 = 0$ and 
$P_1^2 = P_2^2 = L_1^2 = L_2^2 = -m_{\infty}^2 \sim -g^2 T^2$, 
with $m_{\infty}^2$ the asymptotic thermal mass (squared) of the quarks.
(Scalars have a thermal mass which is momentum-independent at one-loop order,
while fermions acquire a momentum-dependent thermal dispersion relation.
But for $P \sim T$, the momentum dependence is sub-leading in $g$
and it is sufficient, at leading order, to just retain the
momentum independent asymptotic thermal mass $m_\infty$.
This asymptotic mass is the same as the thermal mass for scalars
provided the scalar quartic coupling is small compared to $g^2$.)

\begin{figure}
\centerline{\epsfxsize=4.8in\epsfbox{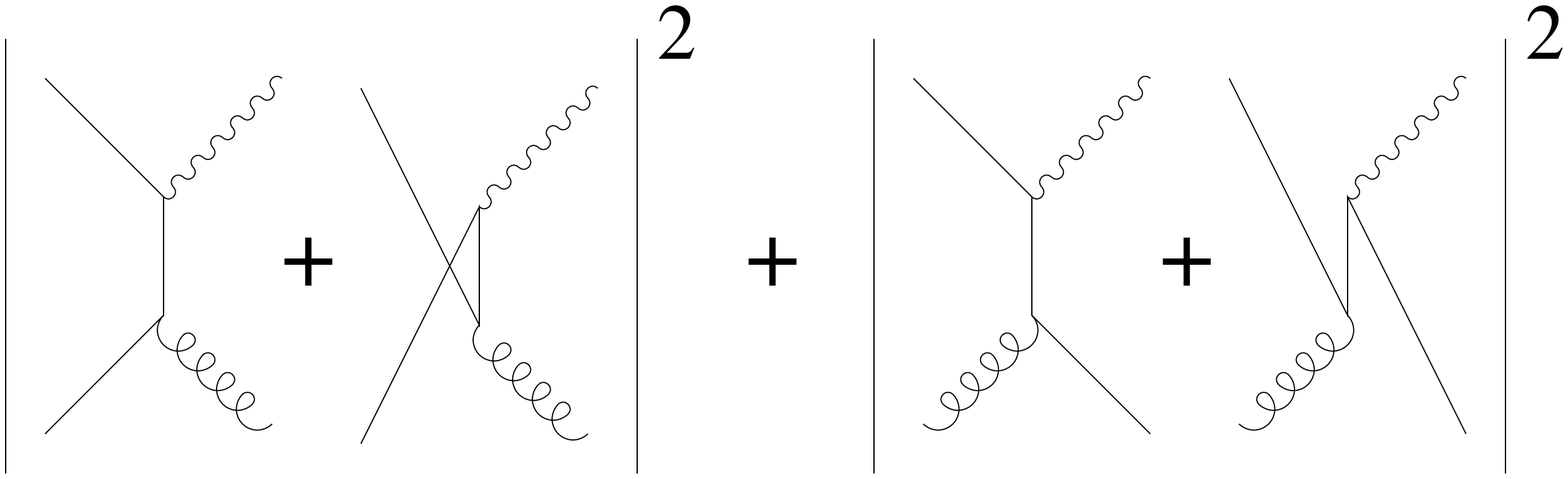}}
\vspace{0.1in}
\caption{\label{fig:2to2} $2\leftrightarrow 2$ 
particle processes which contribute to the
photo-emission rate at $O(e^2 g^2 \, T^4)$.}
\end{figure}

If the exchange momentum $Q$ is of order $T$,
then the rate for this process is $O(e^2 g^4)$, which is negligible
compared to the $2 \leftrightarrow 2$ Compton scattering
and annihilation processes shown in Fig.~\ref{fig:2to2},
whose rates are $O(e^2 g^2)$ up to logs \cite{Kapusta,Baier}.
But for $Q \sim gT$ there is an enhancement from the soft gluon propagator.
The leading contribution will arise from the kinematic regime where
$q^0 \sim |\q| \sim gT$ and $P_1 \cdot Q \sim g^2 T^2$.  
The latter condition restricts the collinearity of the outgoing photon
and quark.
Generically, if all components of $Q$ are $O(gT)$ then
$P_1 \cdot Q \sim gT^2$;
however this region turns out to be subdominant.
(See footnote~\ref {fn:soft} below.)
Because $Q$ is small and the other 4-momenta are almost lightlike,
the condition $Q \cdot P_1 \sim g^2 T^2$,
together with the requirement that $p_2^0$ be at most $O(T)$,
also implies that
$Q \cdot P_2$,
$K \cdot P_2$, and $K \cdot P_1$
are all $O(g^2 T^2)$,
and restricts the spatial components of $\p_1$, $\p_2$, and $\q$
orthogonal to $\k$ to be $O(gT)$.

The emission amplitude for a transverse photon
with polarization vector $\epsilon^\mu$ is
\begin{equation}
\label{eq:amp}
    e \, g^2 \, G^{\mu \nu}_\ret(Q) \,
    (L_2{+}L_1)_\mu
    \left[
    ( P_1 {+} P_2 {+} K)_\nu \,
    \frac{\epsilon \cdot \left( P_1{+}P_2{+}Q \right)}{(P_2+K)^2 + m^2}
    +
    ( P_1 {+} P_2 {-} K)_\nu \,
    \frac{\epsilon \cdot \left( P_1{+}P_2{-}Q \right)}{(P_1-K)^2 + m^2}
    \right] ,
\end{equation}
where the two terms in the bracket come from the diagram where
the photon is radiated from the outgoing $P_2$ line
(shown in Fig.~\ref{intro_fig}) and the corresponding diagram
where the photon is radiated from the incoming $P_1$ line.

One may choose the photon polarization vectors to be
purely spatial in the plasma frame and orthogonal to $\k$.
Therefore the quantities $\epsilon \cdot (P_1{+}P_2{\pm}Q)$ appearing
above are $O(gT)$.
There is no special relation between the two terms which might
cause a cancellation,%
\footnote
    {%
    In the generic soft exchange regime where $Q \cdot P_1 \sim gT^2$,
    the transverse (to $\k$) components of $P_1$ and $P_2$ are
    order $\sqrt g \, T$ and equal [up to $O(gT)$].
    Hence,
    $
	(P_1{+}P_2{\pm}Q)\cdot \epsilon
	\simeq (P_1{+}P_2)\cdot \epsilon
	\sim g^{1/2}T
    $.
    In this case, the two terms in Eq.~(\protect\ref{eq:amp})
    are equal and opposite at leading order and 
    there {\em is\/} a cancellation, related to a cancellation
    addressed by B\"{o}deker in \cite{Bodeker2}.
    This is why the more highly collinear region with
    $Q \cdot P_1 \sim g^2 T^2$ is actually dominant.
    \label {fn:soft}
    }
so it is sufficient to consider the square of one term to estimate the
rate for the process.
There is an explicit $e^2 g^4$ plus a $g^4$ phase space suppression
(from the conditions that all components of $\q$ be
small and that $Q \cdot P_1 \sim g^2 T^2$).
The square of the soft gauge boson propagator 
$G^{\mu \nu}_\ret(Q)$ is of order $1/Q^4 \sim 1/g^4 T^4$, and
the square of the scalar propagator is $1/[(P_2{+}K)^2+m^2_\infty]^2 \sim
1/g^4 T^4$.  Both
$(L_2{+}L_1)_\mu$ and $(P_1{+}P_2{-}K)_\nu$ are $O(T)$.
Finally there are the squares of the numerators,
$[(P_1{+}P_2{+}Q) \cdot \epsilon]^2 \sim g^2 T^2$.
Adding powers of $g$, we find the rate for the process is $O(e^2 g^2 T^4)$,
which is the same order as the rate of the $2 \leftrightarrow 2$ processes
shown in Fig.~\ref{fig:2to2}.
Hence, the inelastic $2 \leftrightarrow 3$ processes of Fig.~\ref{intro_fig}
are of $O(1)$ importance in hard photon emission.%
\footnote
    {%
    Since the incoming and outgoing quark momenta, $P_1$ and $P_2$, 
    differ by $O(T)$, the
    analogous pure QCD processes involving hard gluon emission
    are also of $O(1)$ importance in transport
    coefficients, as recently emphasized in \cite{largeNf}.
    }

Several points about the collinear emission, soft exchange regime
which dominates these $2 \leftrightarrow 3$ processes should be emphasized:
\begin{enumerate}
\item
    The large contribution from a nearly on-shell quark propagator,
    $1/(P_2+K)^2 \sim 1 / g^2 T^2$, is essential.
\item
    The large contribution from a gauge boson propagator with soft
    exchange momentum,
    $1/Q^2 \sim 1/g^2 T^2$, is essential.
\item
    The condition $P_1 \cdot Q \sim g^2 T^2$,
    combined with near-lightlike on-shell incoming and outgoing momenta,
    forces $P_2 \cdot K \sim g^2 T^2$,
    which implies that
    the outgoing photon and quark are highly collinear.
    The opening angle between the outgoing quark and photon 
    satisfies $1-\cos(\theta) \sim g^2$,
    or $\theta = O(g)$.
\item
    Although the above estimates used scalar charge carriers for simplicity,
    all of the conclusions are equally applicable to fermions.
\end{enumerate}

However,
carrying out the calculation of the $2 \leftrightarrow 3$ process as
we have just outlined is not sufficient to obtain the correct
leading order emission rate.
To see this, consider the first point above.
The real part of the inverse scalar propagator
appearing in our (sketch of a) calculation is $O(g^2 T^2)$.
But hard, charged particle self-energies also have $O(g^2 T^2)$
imaginary parts.  
Therefore it appears necessary to include this imaginary part
in the propagator.
But the imaginary part characterizes the probability that the particle
undergoes additional scattering between the time of the initial
scattering shown in the diagram and the time of the photon emission.
If it is necessary to include the imaginary part in an intermediate
propagator,
then it is presumably also important to include such scatterings explicitly,
meaning that we must consider $3 \leftrightarrow 4$ processes,
$4 \leftrightarrow 5$ processes, or more.

\begin{figure}
\centerline{\epsfxsize=4.4in\epsfbox{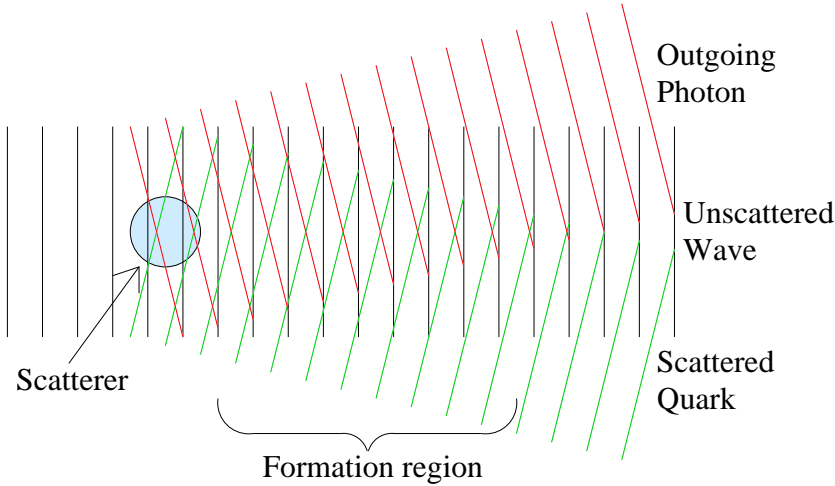}}
\vspace{0.1in}
\caption{\label{fig:formation} A cartoon of the real-space appearance of
scattering and photon emission.  The photon emission,
which is sensitive to the interference of unscattered and scattered waves,
occurs over a region of spatial extent $1/g^2 T$,
which is the same as the mean free path for additional scatterings
of the quark.}
\end{figure}

Alternately, consider point 3~above.
The collinearity condition requires that the components of
the spatial vectors $\p_2$ and $\k$ which are
transverse to $\p_1$ be at most $O(gT)$.
To resolve momenta with that precision, the wave
packets for the outgoing states must have spatial extent of order $1/gT$ in
the transverse direction.  Since the opening angle is order $g$, the
outgoing quark and photon must overlap in 
space over a time interval of order $1/g^2 T$.
This is the formation region; the photon is generated from coherent
interaction between the incoming and scattered waves over a region of
this length, as roughly depicted in Fig.~\ref{fig:formation}.
However, the mean free path for soft scatterings is $O(1/g^2 T)$,
which is the same size.  Therefore
it will be necessary to consider corrections to the emission process
due to scatterings occurring before the photon formation is complete.
Such scatterings can potentially disrupt the photon emission process;
this is called the Landau-Pomeranchuk-Migdal effect \cite{LP1,LP2,M1,M2}.

Because the emission time is $O(1/g^2 T)$, only scattering processes
with a mean free path of order $1/g^2 T$ will be important.  
Therefore we can neglect scattering processes with exchange momentum
parametrically large compared to $gT$.
We will argue that we can also neglect 
any scattering process which changes the direction of the
outgoing quark by an angle small compared to $g$.
In particular, scattering off the $g^2 T$ ultrasoft
gauge boson background will not be important.
This is because the $O(g^2 T)$ momentum 
exchanged in such a scattering is too small to affect the
kinematics of the photon emission process.
It is true that the scattered and
unscattered waves also pick up an $O(1)$ color rotation
(or a phase in an Abelian theory)
from moving through the ultrasoft field over a distance $1/g^2 T$,
but because they are overlapping and collinear,
the color rotations they receive
are the same to within $O(g)$,
and so can not produce an $O(1)$ phase difference which would
affect the photon emission amplitude.
We will see how this physical argument arises in a diagrammatic
analysis in subsection \ref{sec:softer}.

\begin{figure}[t]
\centerline{\epsfxsize=5.0in\epsfbox{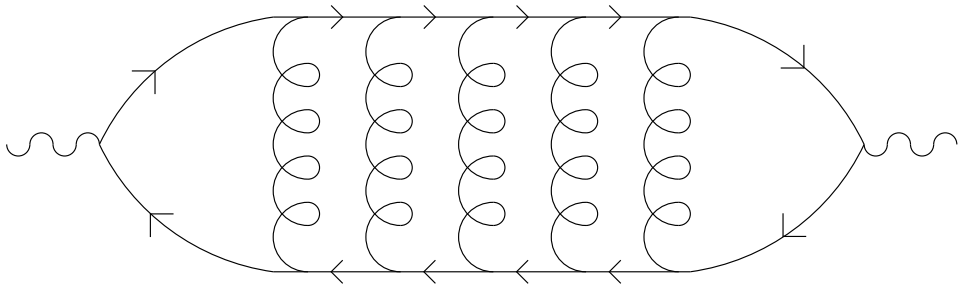}}
\vspace{0.1in}
\caption{\label{fig:ladder} The generic diagram which will contribute to
bremsstrahlung and \ipa\ at leading order.
The solid outer lines represent charged particles whose momenta
are hard but collinear with the (hard, lightlike) incoming photon;
These propagators can all be approximately on-shell simultaneously.
Therefore, self-energy resummation, including the imaginary part of
the self-energy, is required.
The cross-rungs represent soft gauge boson exchange
with momenta $Q^2 \sim g^2 T^2$ which are restricted to satisfy
$Q \cdot K \sim g^2 T^2$ in order to maintain the collinearity condition.}
\end{figure}

In this paper we show how to sum all processes like
Fig.~\ref{intro_fig} but with an arbitrary number of additional
scatterings, working at all times to leading order in $g$.
We show that the important exchange momenta are parametrically $O(gT)$;
both harder and softer exchange momenta are irrelevant at leading order.
Because the spatial extent of a $gT$ scattering event is $O(1/gT)$,
while the mean free path is $O(1/g^2 T)$,
the scatterings can be considered sequential.
Writing the photon emission rate in terms of a
current-current correlator, or equivalently a photon self-energy, 
we show that the dominant set of diagrams
for bremsstrahlung and \ipa\ are precisely ladder diagrams
of the form shown in Fig.~\ref{fig:ladder}.  

\subsection {Results}

Let $\Gamma_\gamma^{\LPM}$ denote the contribution to the photon
emission rate per unit volume,
\begin {equation}
    \Gamma_\gamma \equiv \frac{dn_\gamma}{dV \, dt} \,,
\end {equation}
from bremsstrahlung and inelastic pair annihilation processes.
The contribution $\Gamma_\gamma^{\LPM}$ should be added to the
standard results \cite{Kapusta,Baier}
for $2\leftarrow2$ processes (Fig.~2) to obtain the total
rate at $O(\alpha_{\rm EM} \, \alpha_{\rm s})$.
We assume that the temperature is large compared to
zero temperature masses (or $\Lambda_{\rm QCD}$),
and neglect relative corrections suppressed by powers of $g(T)$
or $m^2/g^2 T^2$.
We also assume that the chemical potential for fermions is at most
of order of the temperature, $\mu \lsim O(T)$, 
but if scalar charge carriers are present, their
chemical potential must be parametrically small compared to $gT$
(since otherwise we would have to deal with Bose condensation
of the scalars --- something we wish to ignore).
Finally, we require that the energy of the photon be parametrically
large compared to $g^4 T \ln g^{-1}$, so that the photon wavelength is
much smaller than the large angle mean free path.

Our result for the (unpolarized) differential emission rate
$d\Gamma_\gamma^{\LPM}$
from a single complex scalar or Dirac fermion species
in an equilibrium plasma at temperature $T$ and chemical potential $\mu$
is given by
\begin{equation}
\label{eq:result1}
    \frac{d\Gamma_\gamma^{\LPM}}{d^3 \k} 
    =
    \frac{\df \,q_{\rm s}^2\, \alphaEM}{4\pi^2 k}
    \int_{-\infty}^{\infty} \frac{d\ppar}{2\pi}
    \int  \frac{d^2 \p_\perp}{(2\pi)^2} \;
    A(\ppar,k) \:
    \Re \, \Big\{ 2 \p_\perp \cdot{\bf f}(\p_\perp;\ppar,k) \Big\} \,.
\end {equation}
Here $\df$ is the dimension of the quark's representation
[$\nc$ in SU($\nc$) gauge theory, or 3 for QCD],
$q_{\rm s}$ is the Abelian charge of the quark
($2/3$ for up type and $-1/3$ for down type),
and $k \equiv |\k|$.
The kinematic factor $A(\ppar,k)$ depends on the 
spin and statistics of the emitting particle,
\begin {equation}
    A(\ppar,k)
    \equiv
    \left\{ \begin{array}{ll}
	\displaystyle 
	\frac{n_b(k{+}\ppar)\, [1{+}n_b(\ppar)]}{2\ppar \, (\ppar{+}k)} \,,
	& \mbox{scalars;}
	\\
	\displaystyle
	\frac{n_f(k{+}\ppar) \, [1{-}n_f(\ppar)]}
	{2[ \ppar \, (\ppar{+}k) ]^2} \,
	    \left[ \ppar^2 + (\ppar{+}k)^2 \right] \, ,
	& \mbox{fermions,}
	\end{array}
    \right.
\end{equation}
with
\begin{equation}
n_b(p) \equiv \frac{1}{\exp[\beta (p {-} \mu)]-1} \,,
\qquad
n_f(p) \equiv \frac{1}{\exp[\beta (p {-}\mu)]+1} \,,
\end{equation}
the standard Bose and Fermi population functions, respectively,
with $\mu$ the chemical potential per quark (not per baryon).

The quantity ${\bf f}(\p_\perp;\ppar,k)$ appearing in the integrand of
\Eq{eq:result1} is the solution to the following linear integral equation
which sums the effects of multiple scatterings occurring during the
photo-emission process,
\begin{eqnarray}
\label{eq:result2}
2\p_\perp & = & i \delta E \; {\bf f}(\p_\perp;\ppar,k) +
	{\pi \over 2} \, \cf \, \gs^2 \, \mD^2 \int
	\frac{d^2 q_\perp}{(2\pi)^2} \,
	\frac{d\qpar}{2\pi} \,
	\frac{d q^0}{2\pi} \>
	2\pi \delta(q^0{-}\qpar)
\nonumber \\ && \hspace{2.0in} {} \times
	\frac{T}{|\q|}
	\left[ \frac{2}{|\q^2 - \PiL(Q)|^2} 
	+ \frac{[1-(q^0/|\q|)^2]^2}{|(q^0)^2 - \q^2 - \PiT(Q)|^2}
	\right]
\nonumber \\ && \hspace{2.0in} {} \times
	\left[ \, \strut
	    {\bf f}(\p_\perp;\ppar,k) - {\bf f}(\q{+}\p_\perp;\ppar,k)
	\right]^{\strut} ,
\end{eqnarray}
with $\cf$ the quadratic Casimir for the quark
[$\cf = (\nc^2{-}1)/2\nc = 4/3$ in QCD],
$\mD$ the leading-order Debye mass,
and $\delta E$ the difference in quasi-particle energies
(defined as real parts of pole positions)
between the state where the photon has been emitted and the state
where it has not,
\begin{equation}
\label{eq:dE_is}
    \delta E \equiv
    k^0 + E_\p\, {\rm sign}(\ppar) - E_{\p{+}\k}\, {\rm sign}(\ppar{+}k) \, .
\end{equation}
For an SU($N$) gauge theory with $\Ns$ complex scalars and $\Nf$ Dirac fermions
in the fundamental representation, the Debye mass is given by \cite {Weldon}
\begin {equation}
   \mD^2 = \frac 16 \, (2 N + \Ns + \Nf) \> g^2 \, T^2 
	+ {\Nf \over 2\pi^2} \> g^2 \mu^2
\end {equation}
%
(where we have allowed a chemical potential $\mu$ for fermionic number but
have taken the chemical potential for scalars to be negligible).
In the denominators of \Eq{eq:result2},
$\PiT(Q)$ and $\PiL(Q)$ are the standard transverse and longitudinal
gauge boson self energies [given explicitly in \Eq{eq:Pi}].

In the final integral (\ref {eq:result1}) for the emission rate,
the region $\ppar > 0$ represents bremsstrahlung emission by a
particle of energy $\simeq k+\ppar$.
In this region, the product of statistical distribution functions
$n(k{+}\ppar) [1{\pm}n(\ppar)]$ contained in $A(\ppar,k)$ 
obviously represent the population function of an incoming quark 
of energy $k{+}\ppar$ times the appropriate
Bose enhancement or Pauli blocking factor
for the outgoing quark of energy $\ppar$.
The region $ -|\k| < \ppar < 0$ represents \ipa.
Using the relations
\begin {equation}
    n_b(-p) = -[1+\barn_b(p)] \,, \qquad
    n_f(-p) = [1-\barn_f(p)] \,,
\end {equation}
with $\barn(p) \equiv 1/[e^{\beta (p+\mu)} \mp 1]$
the appropriate anti-particle distribution function,
the factor $A(\ppar,k)$
in this interval
may be rewritten in the form
\begin {equation}
    A(\ppar,k)
    \equiv
    \left\{ \begin{array}{ll}
	\displaystyle 
	\frac{n_b(k{-}|\ppar|)\, \barn_b(|\ppar|)}
	    {2|\ppar| \, (k{-}|\ppar|)_{\strut}} \,,
	& \mbox{scalars;}
	\\
	\displaystyle
	\frac{n_f(k{-}|\ppar|) \, \barn_f(|\ppar|)}
	    {2[|\ppar| \, (k{-}|\ppar|)]^2} \,
	\left[\ppar^2 + (k{-}|\ppar|)^2 \right] ,\;
	& \mbox{fermions,} 
	\end{array}
    \right.
\end{equation}
which displays the expected population functions for a
particle and anti-particle to annihilate, producing a photon of energy $k$.
Finally, the region $\ppar < -|\k|$ represents bremsstrahlung emission by an
anti-particle of energy $|\ppar|$,
and gives a contribution which is the same as the $\ppar > 0$ region
except for switching $n_{b,f} \to \barn_{b,f}$.

The integral equation
(\ref {eq:result2}) is similar in form to a linearized Boltzmann equation,
but with the convective derivative replaced by
$i\delta E$ which represents a net phase accumulated due
to the energy difference between outgoing and incoming states.
The integral in \Eq {eq:result2} can be interpreted as a
linearized collision integral, with the
${\bf f} (\q{+}\p_\perp;\ppar,k)$ piece representing the gain term
describing the scattering of particles into momentum $\p$,
and the ${\bf f} (\p_\perp;\ppar,k)$ piece the corresponding loss term
describing the scattering of particles out of the mode with momentum $\p$.
An integral equation of this form was first derived by Migdal \cite{M1,M2}
in the context of energy loss of a fast particle traversing ordinary matter.

The explicit form of the energy $E_\p$ of a hard quark with momentum $|\p|$
is given by
\begin{equation}
    E_\p = \sqrt{\p^2 + m_\infty^2} \simeq |\p| + \frac{m_\infty^2}{2|\p|} 
	\simeq |\ppar| + \frac{\p_\perp^2 + m_\infty^2}
	{2 |\ppar|} \,,
\label {eq:Ep}
\end{equation}
where the asymptotic thermal ``mass''
\begin {equation}
	m_\infty^2 = \frac{\cf \, g^2 \, T^2}{4} \, .
\end {equation}
(For scalar quarks with non-negligible quartic coupling,
an additional term $\sim \lambda \, T^2$ also contributes.)
As noted earlier, this asymptotic mass
is the same for a fermion or a scalar,
and the above approximation to the momentum dependence of the quasi-particle
energy is, for hard excitations, accurate to $O(g^2 T)$
which is sufficient for our leading order analysis.
The final form of $E_\p$ in \Eq{eq:Ep} uses an
expansion in $\p_\perp^2 \ll \ppar^2$,
which is justified because it is the region $\p_\perp^2 = O(g^2 T^2)$
which gives the leading (in $g$) contribution to the emission rate;
this expansion was already used in deriving the integral equation
(\ref{eq:result2}).
Substituting the explicit form of $E_\p$ into the definition
(\ref {eq:dE_is}) gives
\begin{equation}
\label{eq:deltaE}
\delta E = \left[ \frac{\p_\perp^2 + m_\infty^2}{2} \right]
	\left[ \frac{k}{\ppar ( k{+}\ppar)} \right]
	,
\end{equation}
for dispersion free,%
\footnote
    {%
    As noted earlier,
    the photon dispersion relation is also modified by
    medium effects, but the correction,
    $k_0^2 - k^2 \sim e^2 T^2$,
    is smaller than the quark dispersion correction by 
    $O(\alphaEM/\alphas) \ll 1$.
    But if we were interested in the emissivity of an electromagnetic plasma,
    then photon dispersion corrections would need to be retained
    when evaluating $\delta E$.
    }
on-shell photon emission,
so that $k_0=|\k|$.
For off-shell photon emission
(which can be relevant for lepton pair production), 
one must add to this expression the off-shell frequency shift $k^0 {-} k$.
All of our analysis, including the justification that
multiple scattering effects must be summed to determine correctly
the leading photon emission rate,
remains applicable to off-shell photons whose four-momenta satisfy
$|k^0 {-} k| \sim g^2 T$.  When $g^2 T \ll |k^0 {-} k| \ll T$, 
our treatment is correct but unnecessarily complicated; scatterings can
be ignored, as either pair annihilation (for $k^0 > k$) or
bremsstrahlung (for $k > k^0$) is kinematically allowed.  For $|k^0 {-} k|
\gsim T$, our treatment breaks down.

The purpose of this paper is to present a derivation of
Eqs.~(\ref {eq:result1}) and (\ref {eq:result2}),
based on a careful diagrammatic analysis showing that
the ladder diagrams of Fig.~\ref {fig:ladder}, and only these diagrams,
contribute to the leading order photo-emission rate.
We begin with a somewhat more technical overview of the problem
in Sec.~\ref{sec:overview}.
Then we review our conventions and the diagrammatic analysis
techniques we will need in Sec.~\ref{sec:conventions}.
The power counting analysis for the hard photon emission rate,
leading to the identification of the
diagrams shown in Fig.~\ref{fig:ladder}, is presented in Sec.~\ref{sec:power}.
This section also shows that the photo-emission rate is {\em not\/}
sensitive to ultrasoft $g^2 T$ scale physics, but only because of
a cancellation between multiple diagrams containing ultrasoft
gauge boson exchanges.
The leading order ladder diagrams are summed in Sec.~\ref{sec:resum}.
These two sections are the crux of the paper.
We discuss various consistency checks, such as confirming
that our approximations give a transverse current-current
correlator and a finite photo-emission rate, in Sec.~\ref{sec:discuss}.

For simplicity of presentation, the bulk of our analysis will be carried
out using scalar quarks instead of fermions, and will focus on the case of
hard, nearly on-shell photons; that is, photon momenta
$\k = O(T)$ with $|K^2| \lsim g^2 T^2$. 
In the final section~\ref{sec:extensions},
we generalize the treatment to the case of fermions,
and discuss the applicability of our analysis to softer
and/or more highly off-shell photons.
For on-shell photons,
we find that the above results for the leading order photon emission
rate remain valid provided the photon momentum is large compared
to the large angle scattering rate, $|\k| \gg g^4T \ln g^{-1}$.

\section{Diagrammatic approach: Overview}
\label{sec:overview}

The differential unpolarized photon emission rate (per unit volume),
at leading order in $e^2$, is given by the well-known relation
\begin {equation}
    d\Gamma_\gamma = {d^3\k \over (2\pi)^3 \, 2|\k|} \>
    \sum_{a=1,2} \epsilon^\mu_{(a)}(\k) \, \epsilon^\nu_{(a)}(\k) \,
    W_{\mu\nu}(K) \,,
\label {eq:dGamma}
\end{equation}
where $K$ is the null photon 4-momentum with 3-momentum $\k$
and positive energy $k^0=|\k|\equiv k$, 
and $W_{\mu\nu}(K)$ is the Wightman electromagnetic current-current
correlator, 
\begin {equation}
    W_{\mu\nu}(K) = \int d^4x \> e^{-i K x} \, 
    \left\langle j_\mu(x) j_\nu(0) \right\rangle .
\label {eq:Wmunu}
\end {equation}
[We use a metric with $({-}{+}{+}{+})$ signature.]
As always, $\langle \cdots \rangle$ denotes an expectation value
in whatever density matrix is of interest,
which in our case is a thermal ensemble
describing the equilibrium plasma.

The photon polarization basis vectors $\{ \epsilon^\mu_{(a)}(\k) \}$
may be chosen to be unit spatial vectors orthogonal to $\k$.
Of course, gauge invariance ensures that the correlator is transverse, 
$K^\mu \, W_{\mu\nu}(K) = 0$,
which implies that one may replace the sum of projections onto
photon polarizations by $g^{\mu\nu}$.
However, for our analysis it will actually be more convenient
to retain explicit transverse polarizations as in (\ref {eq:dGamma}).

\begin{figure}[t]
\centerline{\epsfxsize=6in\epsfbox{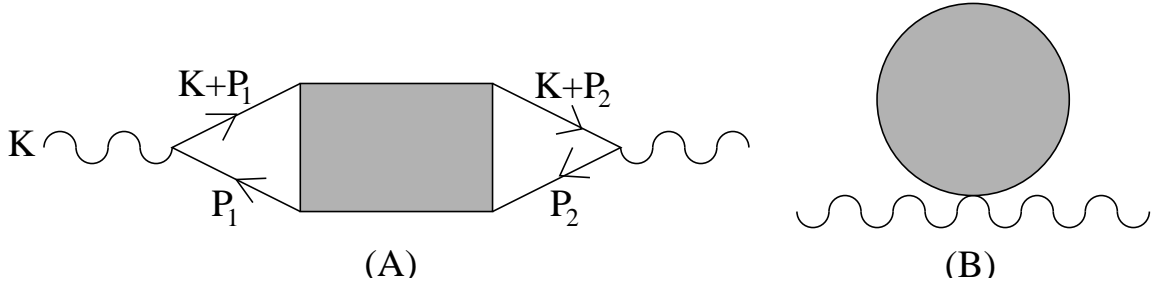}}
\vspace{0.1in}
\caption
    {
    Diagrams contributing to the current-current correlation function.
    The shaded blob in (A) represents the amputated four-point
    correlation function of the electrically charged
    fields which generate the current.
    Tadpole diagrams (B) make no contribution to the
    correlator of interest and may be ignored.
    \label{fig1}
    }
\end{figure}

Note that the product of currents in the expectation value (\ref {eq:Wmunu})
is not time ordered, but is in the order shown.
In other words, we need a Wightman current-current correlation function.
In thermal equilibrium, this is proportional to the spectral density,
or equivalently to the imaginary part of the retarded correlator,
\begin {equation}
    W_{\mu\nu}(K) = 2 \, [n_b(k^0) + 1] \, \Im \, D^\ret_{\mu\nu}(K) \,,
\label {eq:W-spectral}
\end {equation}
with
\begin {equation}
    D^\ret_{\mu\nu}(K) = i \int d^4x \> e^{-iKx} \,
    \Theta(x^0) \,
    \left\langle \left[j_\mu(x), j_\nu(0)\right] \right\rangle ,
\label {eq:D^R}
\end {equation}
and $\Theta(x)$ the usual unit step function.
[Since the current $j_\mu$ is charge neutral,
the Bose distribution function in \Eq{eq:W-spectral}
does not include any chemical potential.]
Because the current $j^\mu(x)$ is bilinear in the fundamental fields,
this correlator is determined by the connected four-point correlation function
of the underlying (electrically) charged fields, as 
shown in Fig.~\ref{fig1}(A).
For theories containing charged scalar fields,
the tadpole contribution to the retarded (or time-ordered) correlator,
shown in Fig.~\ref{fig1}(B),
has no imaginary part.
Hence it does not contribute to the Wightman
function $W_{\mu\nu}$ and may be ignored.

The following discussion will, for simplicity, focus on the case
of a theory with a single scalar ``quark'' with electric charge $e$.
The generalization to fermions appears in Sec.~\ref{sec:fermions}.
We will label momenta as shown in Fig.~\ref{fig1}(A),
so that $P_1$ is the momentum of the line
entering the first current insertion, and $P_2$ 
is the momentum of the line emerging from the second insertion.
Given this, the photo-emission rate per unit volume may
be expressed as
\begin{eqnarray}
d\Gamma_\gamma & = & \frac{\alphaEM}{4\pi^2} \frac{d^3\k}{|\k|}
	\sum_a \epsilon^\mu_{(a)} \epsilon^\nu_{(a)}
	\!\int\! \frac{d^4 P_1}{(2\pi)^4}\frac{d^4 P_2}{(2\pi)^4} \>
	(K{+}2P_1)_\mu (K{+}2P_2)_\nu \>
	G_{1122}(P_1,-K{-}P_1,K{+}P_2,-P_2)
	\nonumber \\
& = & \frac{\alphaEM}{\pi^2} \, \frac{d^3\k}{|\k|}
	\int \frac{d^4 P_1}{(2\pi)^4} \frac{d^4 P_2}{(2\pi)^4} \>
	P_{1,\perp} \cdot P_{2,\perp} \;
	G_{1122}(P_1,-K{-}P_1,K{+}P_2,-P_2)
	\, ,
\label{eq:to_compute}
\end{eqnarray}
where we define $P_\perp$ to be the projection of a 4-momentum $P$
onto the plane orthogonal to both $K$ and $\k$.
In this expression, $G_{1122}$ denotes
the Fourier transform (defined so that all momenta are incoming)
of the scalar field four-point function
\begin {equation}
    G_{1122}(x_1,x_2;y_1,y_2)
    \equiv
    \left\langle \,
	\overline{\cal T} \, \Big\{ \phi^\dagger(y_2) \, \phi(y_1) \, \Big\}
	\: {\cal T} \,\Big\{ \phi^\dagger(x_2) \, \phi(x_1) \, \Big\}
    \right\rangle ,
\label {eq:G1122}
\end {equation}
where ${\cal T}$ denotes time ordering
and $\overline{\cal T}$ anti-time ordering.
For our immediate purposes it is sufficient to note that the lowest order
one loop diagram (where the shaded box is absent and $P_1 = P_2$)
represents the disconnected part of $G_{1122}$ which is just
a product of Wightman correlators,
\begin {equation}
    G^{\rm disc}_{1122}(x_1,x_2;y_1,y_2)
    =
    \left\langle \phi^\dagger(y_2)\,\phi(x_1) \right\rangle
    \left\langle \phi(y_1)\,\phi^\dagger(x_2) \right\rangle \,,
\end {equation}
and that any equilibrium Wightman two-point function (Fourier transformed)
is $n(k^0) {+} 1$ times the corresponding spectral density
[{\em cf.} Eq.~(\ref {eq:W-spectral})].

Throughout our analysis, we consider diagrams in which
self energy insertions on internal lines have been resummed
({\em i.e.}, skeleton diagrams).
For sufficiently high temperature, so that $\alpha_s(T)$ is small,
this means that the spectral density associated with a line
carrying momentum $P$ has a quasiparticle peak at $P^2 = O(g^2T^2)$,
together with a smooth ``off-shell'' background.
The width in energy of the quasiparticle peak is $O(g^2 T \ln g^{-1})$,
and its integrated spectral weight is $O(1)$,
while the spectral weight of the off-shell background is $O(g^2)$%
\cite {damping-rates,Weldon}.
[These estimates assume that the components of $P$ are $O(T)$.
The next section contains a more detailed description.]

To understand which kinematic regimes can contribute to the leading order
hard, on-shell photon emission rate
[{\em i.e.}, $K^2 = 0$ with $|\k| \gsim T$],
it will be helpful to consider separately the following regions.%
\footnote
    {%
    For completeness, there is also the transition region
    in which $P_{1,\perp}$ and/or $P_{2,\perp}$ are $O(g^\nu T)$
    for some power $\nu$ intermediate between zero and one,
    $0 < \nu < 1$.
    Contributions from this regime could potentially give rise to
    logarithms of the form $\ln[T/(gT)] = \ln g^{-1}$.
    But this region never dominates the contributions of both
    the non-collinear and near-collinear regions by a power of $g$.
    Hence, for the purpose of identifying which classes of diagrams
    can contribute to the leading-order photon emission rate,
    this transition region need not be separately considered.
    Nevertheless,
    our final results (\ref {eq:result1}) and (\ref {eq:result2}),
    together with existing treatments of the non-collinear region,
    do correctly include contributions from this intermediate region.
    }
\begin{enumerate}
\item
    Non-collinear photon emission.
    By this, we mean that the perpendicular components of momenta
    $\p_1$ and $\p_2$ (relative to the photon momentum $\k$)
    are both comparable to the temperature,
    so $P_{1,\perp}$ and $P_{2,\perp}$ are each $O(T)$.
\item
    Near-collinear photon emission.
    One or both of the momenta $\p_1$ and $\p_2$ are nearly collinear
    with $\k$. 
    Specifically, $P_{1,\perp}$ and/or $P_{2,\perp}$ are $O(gT)$ or smaller.
\end{enumerate}

\begin{figure}[t]
\centerline{\epsfxsize=5.5in\epsfbox{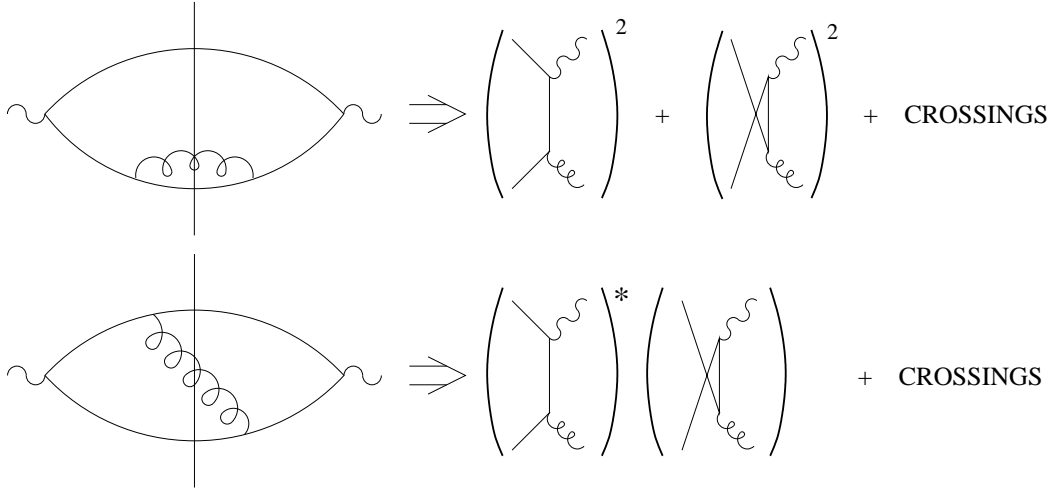}}
\bigskip
\caption{Non-collinear contributions to the current-current correlator,
and their interpretation as products of (quasi-particle) scattering
amplitudes.
The first diagram gives the squares of the amplitudes for individual
scattering processes, while the second diagram generates the appropriate
interference terms.
\label{fig2}}
\end{figure}

The leading contribution from region (1) is comparatively easy to analyze.
For the lowest-order one loop diagram where $P_2 = P_1$,
the condition that $P_{1,\perp}$ be $O(T)$ implies that the
propagators carrying momenta $P_1$ and $P_1{+}K$ cannot
both be nearly on-shell.
The dominant contribution arises when one propagator,
say the one with momentum $P_1$,
gives an on-shell spectral weight,
while the other propagator with momentum $K{+}P_1$
gives an off-shell spectral weight
(or vice-versa).
Such an off-shell spectral weight may be viewed as
a cut through a self-energy bubble, as shown in Fig.~\ref{fig2}.
This may be recognized as describing the lowest-order Compton and
annihilation contributions to the photon production rate.
The appearance of a bubble means that one must also consider other
two loop contributions.
As shown in Fig.~\ref{fig2}, these generate the interference diagrams
required to describe the complete lowest-order Compton and
annihilation processes.
These processes yield an $O(\alphas)$ contribution to the current-current
correlator, since there is always either a loop correction or an $O(g^2)$
off-shell spectral weight involved.
[For fermions, it turns out that there is also a logarithmic infrared
enhancement as the off-shell line becomes soft
({\em i.e.}, parallel and perpendicular components of momentum on
one line both small compared to $T$).]
As these contributions have been computed in detail in previous
literature \cite {Kapusta,Baier}, we will not discuss this piece of
the photon emission rate in any detail.
The contributions from region (1) for higher loop diagrams
are suppressed by additional factors of $\gs$.%
\footnote
    {%
    The power counting analysis of section \ref {sec:power}
    may be easily adapted to verify this.
    As will be discussed in detail below,
    the only way higher loop diagrams could {\em not\/}
    be suppressed is if each additional loop involves either
    soft or collinear enhancements which are sufficient to
    cancel an explicit factor of $g^2$.
    This is precisely what happens in the near-collinear region (2),
    which will be analyzed in detail, but not in the non-collinear region (1).
    In fact, the first corrections to the leading non-collinear
    contributions involve the soft sector and are suppressed by
    one power of $g$.
    }

The collinear region (2) is more difficult to analyze.
The regime where $P_1$ (or $P_2$) is nearly collinear with $K$
is phase space suppressed;
just requiring the momenta to be in the right range
provides a $g^2$ suppression, since both components of $P_{1,\perp}$
must be $O(gT)$.
Moreover, since \Eq{eq:to_compute} contains a factor of
$P_{1,\perp}\cdot P_{2,\perp}$,
there is at least another $O(g)$ suppression.
However, the contour for the frequency integration [over $(p_1)^0$]
is nearly pinched between the on-shell poles of the $P_1$ and $K{+}P_1$
propagators, which leads to a
``pinching pole'' enhancement of the frequency integral
by a factor of the inverse distance between these poles.
These poles are displaced by $\delta E \sim O(g^2 T)$,
which implies
[as discussed in detail in section \ref {sec:power}, {\em c.f.}~\Eq{eq:pinch}]
that the frequency integral generates a $1/g^2$
which compensates the $O(g^2)$ phase space suppression.
Because the precise separation of the on-shell poles of the two propagators
is relevant, the inclusion of correct self-energies on the propagators
is of $O(1)$ importance.
For the lowest-order one loop diagram,
$P_{1,\perp}\cdot P_{2,\perp} = P_{1,\perp}^2 = O(g^2 T^2)$ in this region,
which means the contribution from this collinear region is $O(\alphas)$,
or equally important as the contribution from region (1).
Moreover, higher loop diagrams in this collinear region may contain
additional ``pinching pole'' enhancements which can compensate for the
explicit vertex factors of $g^2$ associated with each additional loop.
This implies that an infinite class of diagrams may contribute to the
leading-order emission rate.
It has been argued \cite{Gelis3}, and we will show below, that all
uncrossed ``ladder'' exchanges of gauge bosons carrying momentum $O(gT)$
contribute at the same order as the simple one-loop diagram.
The next two sections show
that such diagrams, and only such diagrams, are important,
while Sec.~\ref{sec:resum} will show how they may be summed
to obtain \Eqs (\ref{eq:result1}) and (\ref{eq:result2}).

\section {Real-time Thermal Field Theory}
\la{sec:conventions}

The correlation functions we are interested in have the general form
\begin{equation}
    \left\langle
	\overline{\cal T} \, \Big\{ \prod_{j=M+1}^N {\cal O}_j \Big\}
	\: {\cal T} \,\Big\{ \prod_{i=1}^M {\cal O}_{i} \Big\}
    \right\rangle
    \equiv
    \left\langle {\cal T}_C \,
	\Big\{ \prod_{j=M+1}^N {\cal O}_{2,j} \prod_{i=1}^M {\cal O}_{1,i}
	\Big\}
    \right\rangle ,
\label {eq:Kel-corr1}
\end {equation}
where ${\cal T}$ denotes time ordering
and $\overline{\cal T}$ anti-time ordering,
so that, reading from right to left, there are a series of operators
at ascending times followed by a series at descending times.
The scalar field four-point function $G_{1122}$ in
Eq.~(\ref {eq:G1122}) is precisely of this form.
As is well-known, correlation functions of this type may be
represented by functional integrals in which
fields are defined on a complex time contour which doubly traverses
the real time axis forward and back,
and then runs into the lower half plane a distance $-i\beta$ \cite {Keldysh}.
In the second form of Eq.~(\ref {eq:Kel-corr1}),
operators labeled 1 are to be viewed as insertions on the
future-directed part of the contour,
operators labeled 2 are insertions on the past-directed part of the contour,
and ${\cal T}_C$ denotes contour ordering.

All the usual rules of diagrammatic perturbation theory immediately
generalize to this ``closed-time-path'' formalism.
Because the expectation value implicitly contains factors of $\exp(-iHt)$
on the forward-directed branch of the time-contour, and $\exp(iHt)$
on the past-directed branch, interaction vertices on the 1 and 2 branches
have opposite sign.
For example, a $\lambda \phi^4$ interaction produces insertions of
\begin{equation}
    -i \lambda \int d^4x \> \Big( \phi_1^4(x) - \phi_2^4(x) \Big) \, ,
\label{eq:signs}
\end{equation}
where $\int d^4x$ denotes a conventional integral over Minkowski space.%
\footnote
    {%
    Interactions also generate insertions on the ``stub'' of the
    time contour extending into the lower half plane,
    which produce perturbative corrections to the density matrix
    describing the equilibrium thermal ensemble.
    These ``Euclidean'' contributions are not relevant
    to the current discussion.
    }
We will not review the basics for this closed-time-path formalism further,
since it is well treated in several textbooks
\cite{textbook1,textbook2}.

As noted in Refs.~\cite {HW1,HW2},
for calculational purposes it turns out to be quite helpful
to make a change of basis from the upper and lower contour 
fields $\{ \phi_1, \phi_2 \}$
to the so-called $r,a$ basis in which
\begin{equation}
\label{eq:ra}
    \phi_r \equiv \frac{\phi_1 + \phi_2}{2} \, , \qquad
    \phi_a \equiv \phi_1 - \phi_2 \, ,
\end{equation}
and likewise for other fields.%
\footnote
   {%
    This basis was originally proposed by Keldysh \cite{Keldysh}, and
    is sometimes referred to as the Keldysh basis.
    Properties of the $r,a$ basis are discussed for instance
    in Refs.~\cite{Chinese,KobesEijckWeert,HW1}.  The notation is not
    uniform in the literature; another common notation is to
    write our $(1,2)$ as $(+,-)$ and our $(r,a)$ as $(1,2)$.
    }
The reasons for this will be noted shortly.
Propagators in this basis are related to conventional
retarded or advanced propagators, or to the spectral density, as follows.%
\footnote
    {%
    Note that our propagators in the $r,a$ basis are defined without
    any factors of $2$, which differs from the convention of \cite{HW1,HW2}.  
    Also, note that the first $r$ or $a$ index on the propagator refers
    to the field $\phi$, and the second index to $\phi^\dagger$.
    }
\begin {eqnarray}
    G_{rr}(P)
    &\equiv&
    i \int d^4x \> e^{-i P x} \, \left\langle {\cal T}_C
    \left(\phi_r(x) \, \phi_r^\dagger(0)\right) \right\rangle
    =
    i \left[ \half + n_b(p^0) \right] \rho(P) \,,
\label {eq:Grr-def}
\\
    G_{ra}(P)
    &\equiv&
    i \int d^4x \> e^{-i P x} \, \left\langle {\cal T}_C
    \left(\phi_r(x) \, \phi_a^\dagger(0)\right) \right\rangle
    =
    G_\ret(P) \,,
\\
    G_{ar}(P)
    &\equiv&
    i \int d^4x \> e^{-i P x} \, \left\langle {\cal T}_C
    \left(\phi_a(x) \, \phi_r^\dagger(0)\right) \right\rangle
    =
    G_\adv(P) = \left[ G_\ret(P) \right]^* ,
\\
    G_{aa}(P)
    &\equiv&
    i \int d^4x \> e^{-i P x} \, \left\langle {\cal T}_C
    \left(\phi_a(x) \, \phi_a^\dagger(0)\right) \right\rangle
    =
    0 \,.
\end {eqnarray}
Here and throughout we have suppressed writing group indices on
propagators.  
The retarded and advanced propagators are%
\footnote
    {%
    Because $\phi$ is a complex field,
    $G_\ret(-P) \ne G_\adv(P)$, unlike the case of a real field;
    rather $G_\ret(-P)$ is the advanced propagator which differs
    from \Eq{eq:GA-def} by interchanging $\phi$ and $\phi^\dagger$.
    This implies that changing the sign of the four-momentum
    does not convert $G_{ra}$ into $G_{ar}$.
    }
\begin {eqnarray}
    G_\ret(P) &\equiv& i \int d^4x \> e^{-i P x} \,
    \Theta (x^0) \left\langle [\phi(x), \phi^\dagger(0)] \right\rangle ,
\label {eq:GR-def}
\\
    G_\adv(P) &\equiv& -i \int d^4x \> e^{-i P x} \,
    \Theta (-x^0) \left\langle [\phi(x), \phi^\dagger(0)] \right\rangle ,
\label {eq:GA-def}
\end {eqnarray}
while the spectral density
\begin {eqnarray}
    \rho(P) &\equiv& \int d^4x \> e^{-i P x} \,
    \left\langle \left[ \phi(x), \phi^\dagger(0) \right] \right\rangle
\nonumber
\\ &=&
    i \left( G_\adv(P) - G_\ret(P) \right)
    =
    2 \, \Im \, G_\ret(P) \,.
\label {eq:rho-def}
\end {eqnarray}
The spectral density $\rho(P)$ is real, odd in $p^0$, even in $\p$,
and positive for positive $p^0$.
Hence $G_{rr}(P)$
is symmetric in both $\p$ and $p^0$, and
is pure imaginary with positive imaginary part.
Up to a factor of $i$, $G_{rr}$ is the same as
the ordering-averaged two point function,
$
    G_{rr}(P) =
    i\int d^4x \> e^{-i P x} \, \left\langle 
    \half \{ \phi(x) , \phi^\dagger(0)\} \right\rangle
$.

The statistical factor $\half + n_b(p^0) = \half \coth(p^0/2T)$
appearing in relation (\ref {eq:Grr-def}) applies to Bose fields only;
for fermionic fields it is replaced by
$\half - n_f(p^0) = \half \tanh(p^0/2T)$.
Note that both $\half + n_b(p^0)$ and $\half - n_f(p^0)$ are odd functions
of $p^0$.
Also, for fermionic fields the commutators in Eqs.~(\ref {eq:GR-def}),
(\ref {eq:GA-def}) and (\ref {eq:rho-def}) are replaced by anticommutators.
The symmetry and reality properties of the spectral density are the same
for Bose and Fermi fields.

Because interaction vertices are necessarily odd under
the interchange of $\phi_1$ and $\phi_2$, as illustrated by \Eq{eq:signs},
in the $r,a$ basis all vertices must involve an odd number of $a$ indices.
In particular, there are no pure $r$ vertices.
This fact, together with the vanishing of $G_{aa}$, helps make
the $r,a$ basis particularly convenient.
In the following, we will need the explicit form of the
vertex characterizing the scalar-field electromagnetic interaction.
If the scalar lines leading into and out of the vertex carry
momenta $P$ and $P'$, respectively, then the interaction with a single
gauge field is
\begin{eqnarray}
\label{eq:vertex}
    \begin{picture}(35,14)(-15,-2)
	\thicklines
	\put(-10,10){\makebox(0,0)[br]{$P'$}}
	\put(-10,-10){\makebox(0,0)[tr]{$P$}}
	\put(0,0){\vector(-1,2){7}}
	\put(-7,-14){\vector(1,2){7}}
	\put(0,2){\oval(5,4)[br]}
	\put(5,1.5){\oval(5,4)[t]}
	\put(10,2){\oval(5,4)[b]}
	\put(15,1.5){\oval(5,4)[t]}
	\put(20,2){\oval(5,4)[bl]}
    \end{picture}
    &=&
    ig \left[P {+} P' \right]_\mu \left\{ 
	A_1^\mu \, \Phi^\dagger_1 \, \Phi_1
	- A_2^\mu \, \Phi^\dagger_2 \, \Phi_2
	\right\}
\nonumber \\ &=&
    ig \left[P {+} P' \right]_\mu
	\left\{
	    A_r^\mu
	    \left( \Phi^\dagger_r \, \Phi_a + \Phi^\dagger_a \, \Phi_r  \right)
	    + A_a^\mu
	    \left(
		\Phi^\dagger_r \, \Phi_r 
		+ {\textstyle \frac{1}{4}} \, \Phi^\dagger_a \, \Phi_a
	    \right)
	\right\} ,
\end{eqnarray}
where $\Phi$ carries momentum $P$, $\Phi^\dagger$ carries $-P'$,
and $A^\mu$ carries $P'{-}P$.

In the free field limit, a retarded scalar propagator has the standard form
\begin{equation}
    \left. G^{\ret}(P) \right|_{\rm free}
    = \left[ \p^2 + m^2 - (p^0 + i \epsilon)^2 \right]^{-1}
    = \left. \left(G^{\adv}(P) \right)^*\right|_{\rm free} \,.
\end{equation}
In the interacting theory,
a non-trivial self-energy $\Sigma(P)$
will also appear in the denominator.
At finite temperature, the self-energy has a non-zero
imaginary part for all nonzero real $p^0$.
This retarded self-energy is obtained from the
Euclidean self-energy by continuing to real frequencies
from positive imaginary Matsubara frequencies.

For hard momentum and energy, $|\p|, p^0 = O(T)$,
the retarded propagator $G^\ret(P)$
for any basic field (scalars, fermions, or gauge bosons)
has a self-energy whose real and imaginary parts are both $O(g^2 T^2)$.%
\footnote
    {%
    In the case of (massless) fermions,
    whose propagators have the form
    $
	(\nott P + \nott \Sigma)^{-1} =
	(\nott{P}{+}\nott{\Sigma})/(P{+}\Sigma)^2
    $,
    our statements about the size of the ``propagator'' are statements
    about
    $1 / (P{+}\Sigma)^2$,
    and statements about the ``self-energy'' are really estimates for
    the deviation of the rationalized denominator from its free value;
    that is, for $2 P \cdot \Sigma + \Sigma^2$.
    }
This applies both on- and off-shell.%
\footnote
    {%
    The imaginary part of the on-shell self-energy has
    a logarithmic enhancement \cite {damping-rates}
    and is $O(g^2 T^2 \ln g^{-1})$
    due to soft scattering with momentum transfers extending
    from $O(gT)$ down to $O(g^2 T)$, see \Eq{eq:Gamma}.
    Since the goal our power-counting analysis is to classify
    contributions according to powers of $g$, such logarithmic
    effects are not relevant and will be ignored, until we
    reach the discussion of ultrasoft contributions
    in Sec.~\ref{sec:softer}.
    }
Therefore, the spectral weight $\rho(P)$ is $O(g^2/T^2)$
if the momentum $P$ is far off-shell [$P^2 = O(T^2)$].
The propagating pole of the retarded propagator acquires an
$O(g^2 T)$ imaginary part in frequency space,
so for near on-shell hard momenta [$P = O(T)$ with $P^2 = O(g^2 T^2)$]
the spectral density resembles a Lorentzian whose peak value is
$O(1/g^2 T^2)$ and whose width is $O(g^2 T)$.
And for hard energies,
$G_{rr}(P)$ has the same characteristic size as
the spectral density $\rho(P)$ (either on- or off-shell)
since the Bose (or Fermi) distribution functions are $O(1)$.

For soft momentum and energy, $|\p|, p^0 = O(gT)$,
the real part of the retarded self-energy for scalars,
gauge bosons, and fermions remains $O(g^2 T^2)$.%
\footnote
    {%
    These estimates for soft momenta assume vanishing or negligible
    chemical potential.
    It will turn out that we actually only need the results for soft
    spacelike gauge bosons (which are unaffected by the chemical potential);
    soft charged fields will not play a role in our leading-order analysis.
    }
The imaginary part is $O(g^3 T^2)$, except for gauge
bosons and fermions with spacelike momentum $P$, in which case
Landau damping produces an $O(g^2 T^2)$ imaginary self energy
\cite {HTL1,HTL2,HTL3}.
Note that for soft excitations with energies of order $gT$,
an $O(g^3 T^2)$ on-shell imaginary self-energy implies
that the quasiparticle width is $O(g^2 T)$ (up to logs),
just as it is for hard excitations.

\begin{table}[t]
\centerline{
\tabcolsep 10pt

\begin{tabular}{|c|c|c|c|c|c|}\hline
&\multicolumn{2}{|c|}{Hard momentum}
&\multicolumn{3}{|c|}{Soft momentum}
\\\cline{2-6}%
Quantity
&off-shell
&on-shell\vphantom{\Bigg|}%
&\tabcolsep 0pt
\begin{tabular}{c}spacelike\\gauge boson\end{tabular}
&\tabcolsep 0pt
\begin{tabular}{c}on-shell\\boson\end{tabular}
&\tabcolsep 0pt
\begin{tabular}{c}other\\boson\end{tabular}
\\\cline{1-6}%
$G_\ret(P)$ 
& $1/T^2$   & $1/(g^2 T^2)$ & $1/(g^2 T^2)$ & $1/(g^3 T^2)$ & $1/(g^2 T^2)$ \\
$\rho(P)$
& $g^2/T^2$ & $1/(g^2 T^2)$ & $1/(g^2 T^2)$ & $1/(g^3 T^2)$ & $1/(g\, T^2)$ \\
$G_{rr}(P)$
& $g^2/T^2$ & $1/(g^2 T^2)$ & $1/(g^3 T^2)$ & $1/(g^4 T^2)$ & $1/(g^2 T^2)$ \\
\hline
\end{tabular}}
\bigskip
\caption
    {
    \label{table:propagators}
    Characteristic sizes of various propagators in different momentum regions.
    ``Hard, off-shell'' means that $|\p|,p^0 = O(T)$ with $P^2 = O(T^2)$,
    ``Hard, on-shell'' means $|\p|,p^0 = O(T)$ with $P^2 = O(g^2T^2)$,
    and ``Soft'' means $|\p|,p^0 = O(gT)$.
    For hard momenta, the estimates shown apply to scalar, fermion,
    or gauge boson propagators.
    For soft momenta, ``on-shell'' means that the energy is within
    $O(g^2 T)$ of the appropriate scalar or gauge boson quasiparticle peak,
    and ``other boson'' denotes either a scalar propagator with arbitrary soft
    4-momentum $P$, or a gauge boson with soft timelike $P$,
    provided $P$ is outside the respective on-shell regions.
    For soft fermions, all results are the same as for soft gauge bosons
    except that $G_{rr}$ is smaller by a factor of $g$, as there is no $1/g$
    enhancement from the Fermi distribution function.
    The estimates for soft scalars assume negligible
    chemical potential.
    }
\end{table}

Consequently, for generic soft momenta (and energy),
the retarded propagator for any excitation is $O(1/g^2 T^2)$.
For spacelike gauge bosons, the corresponding spectral density
is $O(1/g^2 T^2)$, while $G_{rr}(P)$ is $O(1/g^3 T^2)$
due to the $T/p^0 \sim 1/g$ enhancement from the Bose distribution function.
For soft scalars or timelike gauge bosons, the additional factor of $g$
in the $O(g^3 T^2)$ imaginary self-energy implies that
the spectral density is only $O(1/g T^2)$
so $G_{rr}(P)$ is $O(1/g^2 T^2)$.
These estimates hold provided the energy is off-shell
(relative to the pole position defined by the real part of the self-energy)
by $O(gT)$.
If the energy is within $O(g^2 T)$ of being on-shell, 
then the nearly on-shell retarded propagator and the spectral
density are both $O(1/g^3 T^2)$,
and $G_{rr}$ for scalars or gauge bosons is $O(1/g^4 T^2)$.
The above estimates are summarized in Table~\ref{table:propagators}.

\section {Power counting}
\la{sec:power}

\begin{figure}[ht]
\centerline{\epsfxsize=2.8in\epsfbox{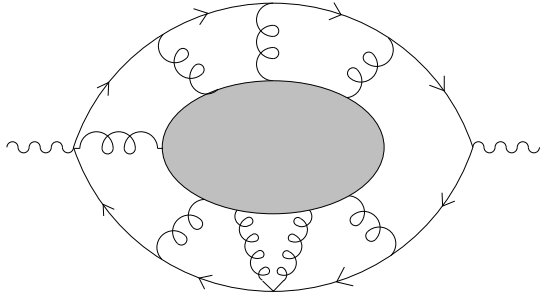}}
\vspace{0.1in}
\caption{
    \label{fig:some_diagrams}
    The important class of diagrams,
    in which both photon lines connect to the same quark loop.
    The blob in the middle connects gluon lines in an arbitrary fashion.
    It need not be connected, and may contain arbitrarily many non-Abelian
    vertices, additional quark loops, and gluon propagators.
    } 
\end{figure}

\subsection{Presence of pinching poles}

\begin{figure}[t]
\centerline{\epsfxsize=3.0in\epsfbox{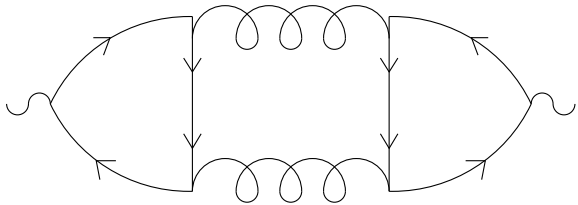}}
\vspace{0.1in}
\caption{
    \label{fig:other_diagrams}
    Simplest non-vanishing 
    diagram in which the photon lines connect to different quark loops,
    which are themselves connected by two or more gluon lines.
    This class of diagrams will be seen to be sub-leading compared
    to the leading diagrams where both photons connect to the same
    quark loop.
    If the quark chemical potential vanishes, then charge conjugation
    invariance implies that there must be three or more gluons
    connecting the two quark loops.
    } 
\end{figure}

We want to evaluate the contribution to the photon emission rate
(\ref{eq:to_compute}) arising from the kinematic region
$P_{1,\perp} = O(gT)$.
Diagrams which can contribute at leading order in the electromagnetic
coupling $e$, but any order in $\gs$,
include those of the general form shown in Fig.~\ref{fig:some_diagrams}
in which both photons connect to the same quark loop.
Any number of gluon lines, which are themselves connected in an
arbitrary fashion, may also attach to the quark loop.
(Recall, however, that all propagators are to be regarded as
containing full self-energies; only skeleton diagrams need be considered
in this discussion, except for a portion of section \ref{sec:softer} below.)
In addition to these diagrams, there are also diagrams in which
the two photon lines connect to different quark loops which are
themselves connected by two or more gluon lines, as illustrated in
Fig.~\ref {fig:other_diagrams}.%
\footnote
    {%
    A single gluon connecting the two quark loops is forbidden
    by (non-Abelian) gauge invariance.
    If the chemical potential of the quark is negligible,
    then charge conjugation invariance implies that there must be
    three or more gluons connecting the two quark loops; and if in
    addition the gauge group is real or pseudo-real (for instance, SU(2)), 
    then a loop with one photon and any number of gluons vanishes.

    At higher orders in $\alphaEM$, diagrams with multiple quark loops
    separated by one or more photon lines are also possible.
    One-particle reducible diagrams with single photons connecting
    quark loops represent re-absorption corrections;
    they can be easily included by interpreting our
    result as a computation of emissivity rather than the emission rate.
    For simplicity, we neglect all higher order in $\alphaEM$ corrections.
    }
The bulk of our discussion will focus on the first category of diagrams,
but we will return to this second category in the next subsection.

\begin{figure}[t]
\centerline{\epsfxsize=2.8in\epsfbox{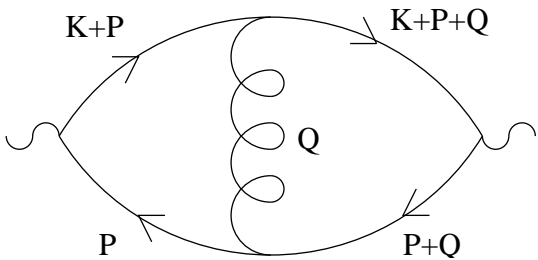}}
\caption{Single gluon exchange diagram.
\label{fig3}}
\end{figure}

It will turn out that a restricted but infinite class of higher loop
diagrams contribute at the same order as the lowest order one-loop graph.
One may understand why as follows.
First, consider a single gluon exchange, as shown in Fig.~\ref{fig3}.
As noted earlier [{\em cf.} Fig.~\ref{fig2}],
this diagram generates the interference part of the leading
$O(\alphas)$ non-collinear contribution.
But focus instead on the near-collinear region where
$P_{\perp}$ is $O(gT)$ and the
exchanged gluon momentum $Q$ is soft,
so that $q^0 \sim gT$ and $|\q| \sim gT$ (and $P{+}Q$ is also 
nearly collinear).
We wish to compare the contribution from this near-collinear region
to the $O(\alphas)$ contribution from the non-collinear region.

For exactly the same reasons discussed at the end of Section
\ref {sec:overview}, the $dp^0$ frequency integration is
dominated by the nearly pinching poles of the $P$ and $K{+}P$
propagators which happens when $P$ is almost null, $P^2 = O(g^2 T^2)$,
and nearly collinear with $K$, so that $K{+}P$ is also almost null,
$(K{+}P)^2 = O(g^2 T^2)$.
Exactly the same reasoning implies that
the dominant contribution to the $dq^0$ frequency integral will occur when
the poles of the $P{+}Q$ and $K{+}P{+}Q$ propagators nearly pinch,
which happens when $(P{+}Q)^2$ and $(K{+}P{+}Q)^2$ are both $O(g^2T^2)$.
These conditions are equivalent to the statement that $Q^2$,
$Q \cdot K$, and $Q \cdot P$ are all $O(g^2 T^2)$.
Now, the gauge boson propagator $G^{\mu \nu}(Q)$ has its indices contracted
against the two $O(T)$ momenta $(2K{+}2P{+}Q)$ and $(2P{+}Q)$,
which we have just argued are approximately lightlike and collinear.
If $G^{\mu \nu}$ were the free propagator, it would behave like
$g^{\mu \nu}$ and hence produce a $g^2$ suppression from contracting
nearly collinear and lightlike 4-momenta.
However, for a soft exchange momentum $Q \sim gT$, the thermal self-energy 
in the gauge boson propagator represents an $O(1)$ modification.
Moreover, the transverse and longitudinal self-energies differ significantly,
which means that the time and space components of the 4-momenta which
$G^{\mu \nu}$ contracts contribute differently.
Hence, there is no such near cancellation.
As noted in the previous section,
$G^{\mu \nu}_{rr} \sim 1/g^3T^2$ for soft spacelike momenta.
Therefore, the gluon propagator contracted between
$(2K{+}2P{+}Q)$ and $(2P{+}Q)$ can be $O(1/g^3)$.

In the near-collinear region, there is a $g^3$ phase space suppression
from the soft $d^3\q$ integration, and a $g^2$ phase space suppression
from the requirement that $P_\perp$ be $O(gT)$
(so that $\p$ is nearly collinear with $\k$).
Finally there is the explicit $g^2$ from gluon exchange vertices,
and a $g^2$ from the $P_\perp \cdot (P{+}Q)_\perp$ factor associated
with the photon vertices.
Combining the pieces, one has a $g^2 \times g^3$
phase space suppression, $g^4$ from vertex factors,
two pinching-pole enhancement factors contributing $g^{-2} \times g^{-2}$,
and a $g^{-3}$ soft gluon propagator.
The net result is an $O(g^2)$ contribution,
exactly the same as the leading-order non-collinear contribution.

Moreover, since the momentum $P{+}Q$ satisfies the same
collinearity conditions as $P$,
adding additional soft gauge boson ``cross-rungs'' will also
give relative $O(1)$ corrections
as each additional rung adds an explicit $g^2 \times g^3$
suppression from the new vertices plus soft phase space for the
new exchange momentum, which is just balanced by a $g^{-2} \times g^{-3}$
enhancement from one more pinching-pole frequency integral
plus another soft gluon propagator.
This will be shown in greater detail in the next section.
Hence, at a minimum, all ladder diagrams will contribute
to the leading order emission rate.

Physically, these gauge boson exchange corrections are important
because the lowest order one loop diagram is dominated by a
narrow range of frequencies which is $O(g^2 T)$ in width.  
That means that the time scale for the emission process is
$\sim 1/g^2 T$, which is the same (up to logarithms) as the
mean free path for small-angle scattering.
Hence, $O(1)$ effects from such scatterings are to be expected.

A similar need to resum ladder diagrams is also encountered when computing
transport coefficients in many other contexts
including the textbook case of conductivity in ordinary metals
(see, for example, Ref.~\cite{Edwards} and Sec.\ 39.2 of Ref.~\cite{Abrikosov}),
viscosity in relativistic scalar field theory \cite{Jeon},
or low frequency electromagnetic response in QCD \cite{LebedevSmilga}.
In all these applications, the presence of nearly
pinching pairs of poles also necessitates the summation of an infinite set
of ladder-like diagrams.  However, because we are considering
hard, lightlike external momentum,
rather than very soft, the appropriate power counting analysis differs
in detail, and cannot simply be taken over from previous work.  
Although the relevant diagram topologies which must be summed
are superficially the same, the kinematic
regions of the diagrams which are important at leading order prove to be
rather different.  In particular, we will
find that only soft momentum flowing through the cross-rungs
will contribute in our application of photo-emission;
this is not true for transport coefficients
in either scalar field theory or gauge theories.

\subsection{Absence of non-Abelian vertices and multiple quark loops}

For ``nice'' problems in thermal field theory,
in which $T$ and $gT$ are the only relevant momentum or frequency scales,
perturbation theory using HTL resummed propagators and vertices
yields an asymptotic expansion in powers of $g$ (not $g^2$),
with each order resulting from only a finite set of diagrams.
This simple diagrammatic structure breaks down whenever
dependence on time scales comparable to the small-angle mean free
scattering time, which scales as $1/(g^2 T)$ (up to logs),
becomes relevant.
It also breaks down whenever dependence on gauge field fluctuations
with spatial scales of $1/(g^2 T)$ or larger becomes relevant,
because the statistics (and dynamics) of such fluctuations is
non-perturbative.

The first condition, dependence on $(g^2 T)^{-1}$ time scales,
is the reason why higher loop diagrams can contribute to the
leading-order hard photon emission rate.
As noted earlier, this dependence arises
from the nearly pinching poles in frequency integrals,
which reflect the presence of virtual intermediate states that
propagate for times of order $(g^2 T)^{-1}$.
However, the only higher loop diagrams which can contribute to the
{\em leading-order} emission rate are those in which each additional
loop brings with it one more pinching-pole enhancement factor.
In particular, diagrams in which the ``blob'' in Fig.~\ref{fig:some_diagrams}
contains vertices and loops are suppressed.  
(Recall that self-energy insertions on all lines are implicitly resummed.
We are {\em not\/} saying that self-energy insertions are sub-dominant
on either the quark propagators forming the perimeter of the loop,
or on the soft gluon propagators.)

\begin{figure}
\centerline{\epsfxsize=3.6in\epsfbox{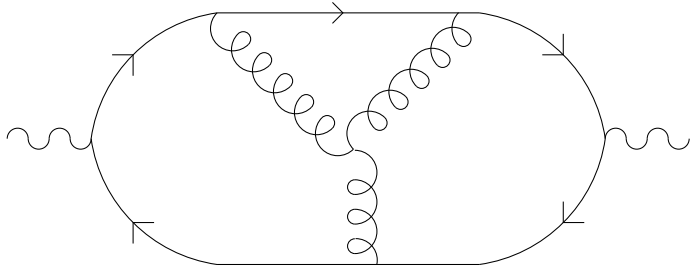}}
\vspace{0.1in}
\caption{\label{fig:threepoint}
    A lowest-order example of a diagram containing a three point non-Abelian
    vertex.
    This, and all other diagrams containing three-point vertices,
    are subleading.}
\end{figure}

As an example,
consider the diagram shown in Fig.~\ref{fig:threepoint}, and compare its
contribution to the one-gluon exchange diagram of Fig.~\ref{fig3}.
Consider the case where all gauge boson lines are soft [$O(gT)$ momenta].
There is an additional $g^2$ suppression from the explicit couplings
on two more gluon vertices and a further $g^3$ suppression from phase space,
from one more soft spatial momentum integral.
Since the three point vertex is a derivative interaction and
all the lines entering it are soft, the vertex contributes an
extra $O(g)$ suppression.
That makes a $g^6$ suppression so far.
Two of the gauge boson propagators can
be $rr$ propagators, but the three point vertex
must have one $a$ index, so one propagator is not.
For soft spacelike momenta, the $rr$ propagators are each
$O(1/g^3 T^3)$, while the soft $ra$ (or $ar$) propagator
is $O(1/g^2 T^2)$;
hence the two additional gauge boson propagators make at most
a $g^{-5}$ contribution.
There is one new frequency integral and
one new scalar propagator.
If the new scalar propagator is on-shell it will
contribute $1/g^2$, but this requires an $O(g^2)$
restriction on the new frequency integration, for a net $g^0$ contribution.
Alternately we could say that one frequency integration involves
a single off-shell scalar propagator and again contributes $g^0$.
Either way, totaling it up, one finds that the diagram is $O(g)$ suppressed.%
\footnote
    {%
    As noted in Table \ref {table:propagators}, 
    the sizes of soft gauge boson propagators are even larger,
    by one power of $1/g$, for on-shell momenta as compared to
    soft spacelike momenta.  But in the frequency region where a
    gauge boson propagator becomes on-shell, a scalar propagator
    leaves the mass shell, so there is no actual enhancement. 
    }
Had the gluon propagators met at a loop rather than a vertex,
the dominant contribution would be from the HTL part of the loop,
which is of the same order as the tree vertex,
so the counting would be unchanged.  

The diagram in Fig.~\ref{fig:threepoint} is even more suppressed
when a hard momentum flows through one or more of the gluons.
Then, except in a phase space restricted range of momentum,
there is not a second pair of pinching poles,
and the diagram is $g^4$ suppressed,
just as the most naive counting would suggest.
The contribution of hard but collinear gluons is also suppressed 
(though by less than $g^4$) for the same reasons discussed in the
next subsection.
The case where the large loop momentum flows through the upper two
gluons, and the upper middle scalar propagator is soft, is also
suppressed, as there is no new pinching pole (see Sec.~\ref{sec:cross}), 
and the gluon propagators $G^{\mu\nu}$ contract nearly null, collinear
momenta, leading to an additional suppression.

More or higher order vertices lead to further suppression,
because they fail to introduce new pinching poles.
For instance, adding another gauge boson line to
Fig.~\ref{fig:threepoint} and replacing the three point vertex
by a four point vertex
brings in one more $g^3$ phase space suppression (in the soft momentum region),
compensated by a $1/g^3$ from the additional soft gauge boson propagator.
The explicit $g^2$ on the four point vertex counts the same as the $g$
of the three point vertex together with the extra $g$ from the derivative
coupling in the three point case.
There is one more explicit $g$ from the new scalar-gauge vertex,
a new frequency integral and one more scalar propagator,
but no additional pinching pole pairs.
Hence, relative to the three point vertex, the four point vertex is suppressed
by one more power of $g$.
Essentially the same arguments rule out diagrams containing a
``seagull'' scalar-gauge field vertex, such as the one illustrated
in Fig.~\ref{fig:some_diagrams}, or diagrams where one of the photons
emerges from a 4 point photo-gluon vertex,
also illustrated by Fig.~\ref{fig:some_diagrams}.

Diagrams of the form illustrated in
Fig.~\ref{fig:other_diagrams} are also suppressed.  
If the gluons are hard and collinear, and the quark ``crossbars'' (the
vertical lines in the figure) are soft, the diagram is suppressed by
$g^6$ with respect to the leading order processes; a factor of $g^2$
arises because the size of $G_{rr}$ in the soft, space-like region is
$g$ smaller for a quark (either scalar or fermionic) than for a gluon,
and a factor of $g^2$ arises from each gluon because they contract
(almost) lightlike collinear momenta (see next subsection).  
The case where the quark crossbars are hard is analogous to the cases
discussed in the next subsection.  Also note that,
in a C invariant theory at zero chemical potential, this diagram
vanishes; three or more gluons must run between quark loops to provide a
nonzero result, and such diagrams are even more suppressed.

\begin{figure}
\centerline{\epsfxsize=4.5in\epsfbox{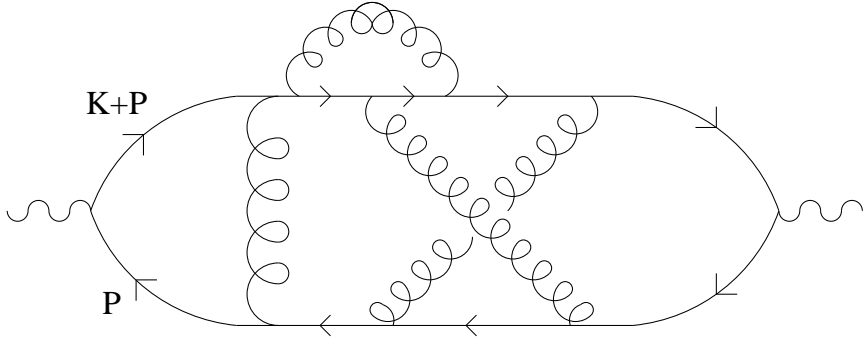}}
\vspace{0.1in}
\caption{An example of the remaining class of diagrams which must be
considered, in which all gauge boson lines connect (at both ends)
to the main quark loop.
\label{fig:no_3pt}}
\end{figure}

Hence, neglecting for the moment the possibility of ``ultrasoft''
gauge boson lines carrying momenta of order $g^2 T$,
the only diagrams we
need consider are those where all gluon propagators attach at both ends
to the ``main'' quark loop.  That is, we need now only consider diagrams like
that shown in Fig.~\ref{fig:no_3pt}.
We will return to ultrasoft momentum bosons
later, as they require a more involved analysis.

\subsection{Momenta harder than $gT$}
\label{sec:hard}

The power counting of hard gluon exchange is relatively simple,
so we will address it next.
Consider Fig.~\ref{fig3} once again, but now when the momentum
$Q$ is hard, $Q \sim T$.
For generic momentum $\q$, the mass shells of $Q$,
$K{+}P{+}Q$, and $P{+}Q$ do not occur at the same value of $q^0$,
so the $q^0$ frequency integral over the three propagators is $O(1)$.
The explicit $g^2$ from the added
vertices therefore suppresses the contribution.
Consider instead the case of collinear momentum $\q$,
that is, $\q_\perp \sim gT$.
In this case, all scalar propagators and the gauge boson propagator can go
on-shell simultaneously.  This kinematic region is $g^2$ suppressed, as
both components of $\q_\perp$ must be $O(gT)$.  It also has the explicit
$g^2$ from the vertices.  At values of the frequency $q^0$ where $Q^2$
is far off shell, there is no enhancement, so this region is suppressed.
The frequency range of $q^0$ which brings all propagators on-shell
is $O(g^2T)$ wide, and there are three new $1/g^2$ propagators,
giving $1/g^4$.
But in this regime, the gauge boson
propagator is contracting near-collinear momenta.
For hard momenta, the Lorentz structure of $G^{\mu \nu}$
is approximately $g^{\mu \nu}$ (in Feynman gauge), but
$g^{\mu \nu} (K{+}2P{+}Q)_\mu (2P{+}Q)_\nu \sim g^2 T^2$.
Therefore, this hard but collinear region is also $g^2$ suppressed.%
\footnote
    {%
    The assumed collinearity conditions also imply that
    $Q \cdot (2P{+}Q)$ and $Q \cdot (K{+}2P{+}Q)$ are $O(g^2 T^2)$.
    Hence, the conclusion that the hard but collinear region is $O(g^2)$
    suppressed does not rely on the particular choice of Feynman gauge.
    }

Physically, the suppression of $O(T)$ scattering momenta comes about
because the time scale involved in the emission process is $1/g^2 T$,
but the mean free path between large angle scatterings is $O(1/g^4 T)$
(up to logarithms).
Hence, the chance of a hard scattering occurring during the
photon emission process is $O(g^2)$.

\subsection{Crossed rungs and soft vertex corrections}
\label{sec:cross}

The next task is to rule out crossed rungs.
Unlike the case of scalar field theory \cite{Jeon},
this is not just a matter of counting the parametric size of each
momentum integral.
Consider the diagram drawn in Fig.~\ref{fig:no_3pt}.  It is
not of the form of the diagrams in Fig.~\ref{fig:ladder}, and we claim
it does not contribute at leading order.  Yet, when all gluon momenta
are $O(gT)$, simple power counting arguments suggest the diagram should
contribute.
Each gauge boson propagator introduces a $g^2$ from its vertices and
a $g^3$ of phase space suppression.
But each gauge boson propagator, for soft spacelike momenta,
has a spectral weight times thermal distribution function which
is $O(1/g^3 T^2)$.
Further, each line has an $O(gT)$ transverse component, so they do not
disturb the collinearity conditions on the hard momenta carried
by the quarks.
In a restricted frequency range, in which each loop frequency is
tuned within an accuracy of $g^2 T$, 
the quark lines are on-shell (within $g^2 T$ in energy)
simultaneously, providing a $1/g^2$ per propagator.
Naively adding powers of $g$, this diagram should
contribute at leading order in $\gs$.
Nevertheless, we will see that when the frequency integrations
are actually carried out, the diagram proves to be suppressed.

The analysis in this subsection is applicable when the momenta
carried by gluons are all much larger than the $g^2 T$ ultrasoft scale;
these arguments will not be applicable to ultrasoft gluons.
However, ultrasoft gluons carry so little momentum that
if any diagram with purely $gT$ scale gluon lines is
``clothed'' by adding additional ultrasoft gluons,
this will not affect any of the arguments of this subsection
concerning $gT$ scale gluons.
Hence it is safe to postpone treating ultrasoft gluons,
and focus first on the analysis of $gT$ scale gluon exchanges.

Crossed diagrams are suppressed, but uncrossed ladders are not,
because of the details of the frequency integrations.
The analysis turns out to be simplest in the $r,a$ basis.
But to use this basis, we must first transform $G_{1122}$ into a
linear combination of $r,a$ basis four-point functions.
By inverting the relation between $r,a$ and $1,2$ bases given in \Eq{eq:ra},
one may easily express $G_{1122}$ as a linear combination of the
16 possible four point $r,a$ correlation functions.
However, not all of the 4-point functions are independent in the $r,a$ basis,
and the relations between them have been worked out recently by Wang and
Heinz \cite{HW1}.  They show that 
$G_{1122}$ can be expressed as a linear
combination of just 7 of the 16 $r,a$ basis four-point functions
and their charge conjugates,%
\footnote
    {%
    Actually, Wang and Heinz \cite {HW1} only discuss the case
    of a real scalar field, while we need the generalization
    to complex fields.
    Using only CPT invariance plus the KMS conditions satisfied
    by thermal equilibrium correlation functions, one may show
    that the appropriate generalization of the results of
    Ref.~\cite {HW1}, for bosonic fields,
    merely replaces the complex conjugate
    $G^*_{\alpha_1 \cdots \alpha_n}$ appearing in Ref.~\cite {HW1}
    by the complex conjugate of the charge conjugated correlator.
    If the equilibrium ensemble is charge conjugation invariant
    (which implies vanishing chemical potentials) then there is no difference.
    }
\begin {eqnarray}
    G_{1122}
    &=&
      \alpha_1 \, G_{aarr} + \alpha_2 \, G_{aaar} + \alpha_3 \, G_{aara} +
      \alpha_4 \, G_{araa} + \alpha_5 \, G_{raaa} + \alpha_6 \, G_{arra} +
      \alpha_7 \, G_{arar}
\nonumber\\ &+&
      \beta_1 \,\bG_{aarr}^{\;*} + \beta_2 \,\bG_{aaar}^{\;*} +
      \beta_3 \,\bG_{aara}^{\;*} + \beta_4 \,\bG_{araa}^{\;*} +
      \beta_5 \,\bG_{raaa}^{\;*} + \beta_6 \,\bG_{arra}^{\;*} +
      \beta_7 \,\bG_{arar}^{\;*} \,,
\label{eq:g1122}
\end {eqnarray}
where all correlation functions
have arguments $({\cal P}_1,{\cal P}_2,{\cal P}_3,{\cal P}_4)$.
Here, 
\begin {equation}
    G_{\lambda_1 \cdots \lambda_4}({\cal P}_1,{\cal P}_2,{\cal P}_3,{\cal P}_4)
    \equiv
    \int \prod_{i=1}^4 d^4 x_i \> e^{-i {\cal P}_i \cdot x_i}
    \left\langle \,
	{\cal T}_C \,
	\Big\{
		\phi^\dagger_{\lambda_4}(x_4) \, \phi_{\lambda_3}(x_3) \, 
		\phi^\dagger_{\lambda_2}(x_2) \, \phi_{\lambda_1}(x_1) \,
	\Big\}
    \right\rangle ,
\end {equation}
while $\bG_{\lambda_1 \cdots \lambda_4}$ denotes the charge conjugated correlator,
\begin {eqnarray}
    \bG_{\lambda_1 \cdots \lambda_4}({\cal P}_1,{\cal P}_2,{\cal P}_3,{\cal P}_4)
    &\equiv&
    \int \prod_{i=1}^4 d^4 x_i \> e^{-i {\cal P}_i \cdot x_i}
    \left\langle \,
	{\cal T}_C \,
	\Big\{
		\phi_{\lambda_4}(x_4) \, \phi^\dagger_{\lambda_3}(x_3) \, 
		\phi_{\lambda_2}(x_2) \, \phi^\dagger_{\lambda_1}(x_1) \,
	\Big\}
    \right\rangle
\nonumber\\ &=&
    G_{\lambda_2\lambda_1\lambda_4\lambda_3}
    ({\cal P}_2,{\cal P}_1,{\cal P}_4,{\cal P}_3)
    \,.
\label{eq:remove_bar}
\end {eqnarray}
If the chemical potentials vanish, so the plasma is charge conjugation
invariant, then
$
    \bG_{\lambda_1 \cdots \lambda_4}({\cal P}_1,{\cal P}_2,{\cal P}_3,{\cal P}_4)
    =
    G_{\lambda_1 \cdots \lambda_4}({\cal P}_1,{\cal P}_2,{\cal P}_3,{\cal P}_4)
$.
With our conventions,%
\footnote
    {%
    Our conventions differ from \pcite{HW1}; their $G$ is
    $i\, 2^{n_{r}-1}$ times our $G$, where $n_r$ is the number
    of $r$ indices on the correlator.
    }
\begin {equation}
\label{eq:alpha}
    \alpha_1 = n[{\cal P}_1] \, n[{\cal P}_2] \,, \qquad
    \beta_1 = - (1{+}n[{\cal P}_3]) \, (1{+}n[{\cal P}_4]) \,
	    \frac{1{+}n[{\cal P}_1]{+}n[{\cal P}_2]}
		 {1{+}n[{\cal P}_3]{+}n[{\cal P}_4]} \, ,
\end {equation}
with $n[{\cal P}] = n_b({\cal P}^0)$
shorthand for the indicated Bose distribution function.
The remaining coefficients $\alpha_2 \ldots \alpha_7$ and
$\beta_2 \ldots \beta_7$
involve similar combinations of distribution functions whose explicit form
can be found in Ref.~\cite{HW1},
but will not be needed for our analysis as
none of these other terms will turn out to contribute at leading order.

The incoming momenta ${\cal P}_1 \cdots {\cal P}_4$ are related
to our previous choice of momentum labels [as shown in Fig.~\ref{fig1}]
through
$
    ({\cal P}_1,{\cal P}_2,{\cal P}_3,{\cal P}_4)
    =
    (P_1,-K{-}P_1,K{+}P_2,-P_2)
$.
Since the momenta $P_1$ and $P_2$ are hard but all exchange momenta are soft,
at leading order
${\cal P}_4 \simeq -{\cal P}_1$ and ${\cal P}_3 \simeq -{\cal P}_2$,
in which case,
\begin {equation}
    \beta_1 \simeq n[{\cal P}_1] \, n[{\cal P}_2] = \alpha_1 \,.
\label{eq:alphabeta}
\end {equation}
Consequently, $G_{aarr}$ and $\bG_{aarr}^{\;*}$ enter with the same
coefficient.  
The contribution of $\bG_{aarr}^{\;*}$ to the emission rate
(\ref {eq:to_compute}) is just the complex conjugate of the
$G_{aarr}$ contribution (even with non-zero chemical potentials),
as may be seen by using the relation (\ref{eq:remove_bar})
to rewrite $\bG_{aarr}(P_1,-K{-}P_1,K{+}P_2,-P_2)$
as $G_{aarr}(-K{-}P_1,P_1,-P_2,K{+}P_2)$,
making a shift $P_1 \to -K{-}P_1$, $P_2 \to -K{-}P_2$
in the integration momenta in \Eq{eq:to_compute},
and noticing that this shift has no effect on the other factors
in the integrand.
As a result, it is only the real part of $G_{aarr}$
which contributes to the emission rate.

\begin{figure}
\centerline{\hspace{1.0in}\epsfxsize=4.8in\epsfbox{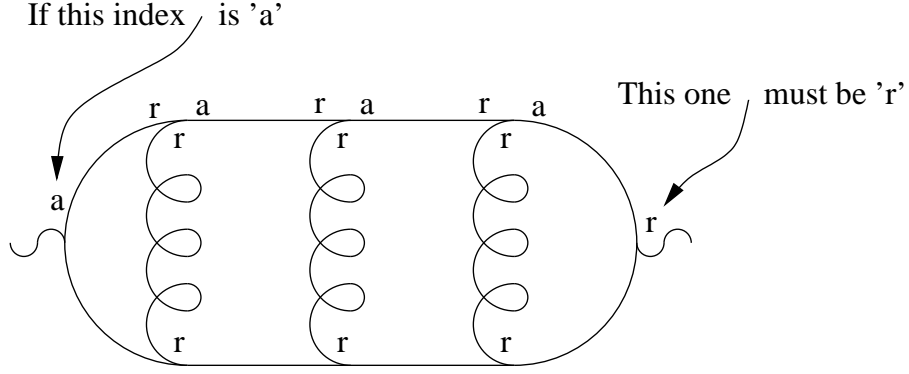}}
\vspace{0.1in}
\caption{\label{fig:ar} Assignment of $r,a$ indices in a ladder diagram.
Because each vertex must have an $a$ index,
the gauge rungs must be $rr$,
and the $aa$ propagator vanishes,
an $a$ index at the beginning of a ladder ``side-rail''
forces alternating $r,a$ assignments for the rest of the rail,
with the other end necessarily ending in $r$.}
\end{figure}

As we have already seen, each new gluon propagator introduces a $g^2$
from vertices and a $g^3$ phase space suppression, as well as a
$1/g^2$ from the pinching-pole frequency integration involving the
two new nearly on-shell scalar propagators.
To contribute at leading order, the gauge boson propagator
must be $O(g^{-3})$, which means it must be an $rr$ propagator
(see Table \ref{table:propagators}).
As shown by \Eq{eq:vertex}, every vertex
must have one $r$ and one $a$ scalar index.
Further, $G_{aa}$ vanishes.
According to \Eq{eq:g1122}, we need only consider diagrams for
four-point functions with at least two $a$ indices.
Consider a propagator emerging from a photon vertex,
with an $a$ assignment at the vertex.
The other end must be an $r$, as $G_{aa}=0$.
Since the gluonic vertex it goes into must have an $a$,
and the gluon uses the other $r$,
the next scalar propagator emerging from the vertex
must begin with an $a$ index.
Regardless of the arrangement of the gauge propagators
(crossed, uncrossed, {\em etc}.),
the scalar line which enters the photon vertex on
the other side has an $a$ index on the gluonic vertex it
leaves, and must therefore enter the photon vertex with
an $r$ index, as illustrated in Fig.~\ref{fig:ar}.
Given the convention for ordering indices on the four-point
function shown in \Eq{eq:G1122}, this demonstrates that only the
$G_{aarr}$ and $G_{arar}$ terms in \Eq{eq:g1122} can contribute
at leading order. 

Return now to Fig.~\ref{fig:no_3pt}.  
Consider the two quark propagators leaving the first photon vertex,
and regard $P$ as the loop momentum circulating around the left-most
loop of the diagram.
Fix for the moment the three spatial components of $P$,
and consider the integration over the frequency $p^0$.
As far as this frequency integral is concerned, the essential
part of the integrand is the product of scalar propagators
with momenta $P$ and $K{+}P$;
all other $p^0$ dependence is sufficiently weak that it may be
neglected over the relevant frequency interval which is only $O(g^2 T)$ wide.
This will be discussed in greater detail momentarily.
If the photon vertex is an $aa$ vertex, then one has
\begin{eqnarray}
    \int \frac{dp^0}{2\pi} \> G_{ar}(K+P) \, G_{ra}(P)
    &=& 
    \int \frac{dp^0}{2\pi} \> G_\adv(K+P) \, G_\ret(P)
\\ &\simeq&
    \int \frac{dp^0}{2\pi} \>
    \frac{1} {\Big[ (p^0+\frac i2\, \Gamma_\p)^2 - E_\p^2 \Big]
	  \Big[ (p^0+k^0-\frac i2\, \Gamma_{\p{+}\k})^2 - E_{\p{+}\k}^2 \Big]} \, .
\nonumber
\end{eqnarray}
Here, $E_\p$ is the on-shell quasiparticle energy,
$
    \Gamma_\p \equiv \Im \Sigma(E_\p,\p) / E_\p
$
is the quasiparticle decay width generated by the imaginary part of the
on-shell self-energy,
and the Lorentzian approximation to the propagators is valid for
frequencies near the quasiparticle poles.
[In thermal field theory $\Gamma_\p/2$ is often called the damping rate.]
The integrand has four poles, two above and two below the real axis.
The pole locations are
\begin {equation}
    p^0 = \pm E_\p - i \Gamma_\p/2 \,, \qquad
    p^0 = -k^0 \pm E_{\p{+}\k} + i \Gamma_{\p{+}\k}/2 \,.
\end {equation}
Since $\p$ and $\k$ are nearly collinear, $E_{\p{+}\k} \simeq E_\p \pm |\k|$
(with the $\pm$ depending on whether $\p$ and $\k$ are aligned or anti-aligned).
In either case, two pole positions almost coincide,
at approximately $p^0 = \ppar \equiv \p \cdot \hat{\k}$.  
One is from the retarded propagator and one is from the advanced propagator,
so they are on opposite sides of the contour.
That is, the poles nearly pinch the integration contour.
The integral is dominated by the region near the pinching poles
and one finds
\begin{equation}
\label{eq:pinch}
    \int \frac{dp^0}{2\pi} \> G_{\adv}(K+P) \, G_{\ret}(P)
    \simeq
    \frac{1}
    {
    4 \ppar \, (\ppar{+}k)
    \left[\half(\Gamma_\p {+} \Gamma_{\p{+}\k}) + i \, \delta E\right]
    } \, ,
\end{equation}
where we have defined
\begin{equation}
    \delta E
    \equiv
    \left[ E_\p\, {\rm sign}(\ppar) - \ppar \right]
	- \left[ E_{\p{+}\k}\, {\rm sign}(\ppar {+} k) - \ppar -k^0 \right ] ,
\label {eq:dEdef2}
\end{equation}
as in \Eq{eq:dE_is},
and approximated $\sgn(\ppar)\, \sgn(\ppar{+}k) \, E_\p \, E_{\p+\k}$
by $\ppar (\ppar{+}k)$ in the denominator of the result (\ref{eq:pinch}),
since the difference is subleading in $g$ in the relevant
domain where $\ppar$ and $\ppar{+}k$ are $O(T)$ while $\p_\perp$ is $O(gT)$.
Note that $\delta E$ and $\Gamma_\p$ are both $O(g^2 T)$
in this domain
[assuming, as we do throughout this discussion, that the photon is
off-shell by at most a comparable amount, $k{-}k^0 = O(g^2 T)$].
Consequently, the width of the frequency interval which provides the
dominant contribution to the integral is of order $g^2 T$.
The factor of $\half(\Gamma_\p {+} \Gamma_{\p{+}\k}) + i \, \delta E$
in the denominator of (\ref {eq:pinch}) is nothing other than
($i$ times) the separation between the nearly pinching poles in the complex
frequency plane.
It should be borne in mind that $\delta E$ is a function of both
$\p$ and the photon four-momentum $K$, even though this is not
indicated explicitly.
For hard momentum, the quasiparticle decay width $\Gamma_\p$ is
essentially constant in $\p$ and differs from the asymptotic value 
$\Gamma \equiv \lim_{\p\to\infty} \Gamma_\p$ only by an amount
which is subleading in $g$.
As this will have no bearing on any of our results,
henceforth we will generally simplify our expressions
by omitting the spatial momentum dependence of $\Gamma_\p$.

The analysis leading to \Eq{eq:pinch}
ignored the frequency dependence of the other parts
of the integrand such as the gluon propagator, and also ignored
frequency dependence in the imaginary part of the self-energy
as one moves away from its on-shell value.
But such frequency dependence has a characteristic scale of $gT$
or larger, and so is negligible over the parametrically small
$O(g^2 T)$ frequency region which dominates the integral.%
\footnote
    {%
    This assertion is actually a bit too cavalier,
    because $\Gamma_\p$ is logarithmically sensitive to ultra-soft
    $g^2 T$ scale physics \cite {damping-rates}, and the
    $g^2 T$ scale contributions to $\Gamma_\p$ do change
    significantly in an $O(g^2 T)$ frequency range about the pole.
    However, we will see in the next subsection that the sensitivity
    to $g^2 T$ scale physics cancels when one sums over all
    ultra-soft gluon exchanges.
    For this subsection, where we are ignoring ultra-soft gauge bosons,
    one should also ignore $g^2 T$ scale contributions to the
    damping rate, and regard $\Gamma_\p$ as defined by an
    infrared cutoff at a scale $\mu$ such that
    $g^2 T \ll \mu \ll gT$.
    }

In contrast, if the photon vertex has one $r$ and one $a$ index,
as occurs for $G_{arar}$,
then the integral we must consider is
\begin{equation}
\int \frac{dp^0}{2\pi} \> G_{\ret}(P+K) \, G_{\ret}(P) \, ,
\end{equation}
and this has the same pole positions,
except that the signs of the imaginary parts are all now the same.
Therefore the contour for the frequency integration may be deformed
away from these poles [by a distance large compared to the $O(g^2 T)$
displacement of the poles from the real axis],
which implies that the integral has no $1/g^2$ enhancement,
unlike the pinching-pole result (\ref{eq:pinch}).

Therefore, of the various different $r,a$ basis correlators appearing
in \Eq{eq:g1122}, only $G_{aarr}$ (and its charge conjugate) contributes
at leading order;
$G_{arar}$ and all the others may be dropped.
This substantial simplification is what makes the $r,a$
representation convenient for this problem.
A similar reduction has been noted recently for
shear viscosity in scalar field theory \cite{HW2}.

\begin{figure}
\centerline{\epsfxsize=2.6in\epsfbox{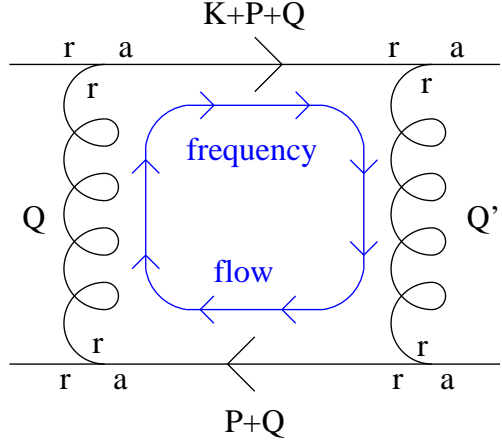}}
\vspace{0.1in}
\caption{\label{fig:uncrossed}
    An individual loop in a ladder diagram with uncrossed rungs,
    with $r,a$ vertex assignments consistent with a leading-order
    pinching-pole contribution.
    }
\end{figure}

Although it is not strictly necessary, it may be helpful
at this point to verify explicitly that ladder diagrams
with uncrossed rungs really do contribute at leading order ---
there is no unforeseen cancellation which would make their
contribution subleading.
Consider the diagram fragment shown in Fig.~\ref{fig:uncrossed},
and integrate over the frequency flowing around the loop as shown.
If both gauge boson propagators carry soft momentum, $Q \sim Q' \sim gT$,
then this is the scale of variation in frequency of the gauge boson
propagators.
And as before, this implies that over the frequency interval of
width $g^2 T$ in which the $P{+}Q$ and $K{+}P{+}Q$ scalar propagators
have nearly pinching poles,
the gauge boson propagators are constant up to $O(g)$ corrections.
Therefore they factor out of the frequency integral,
which as we have already seen contains a pair of poles
spaced $O(g^2 T)$ apart.
So, just as for the end-most loop examined earlier,
we have an integral of form
$\int dq^0 \> G_\adv (K{+}P{+}Q) \, G_\ret (P{+}Q)$,
which is $O(1/g^2 T^3)$.
Therefore, every pair of scalar lines between uncrossed gauge boson rungs
gives rise to a pinched frequency integral which contributes a $1/g^2$,
as required to make uncrossed ladders contribute at leading order.

\begin{figure}
\centerline{\epsfxsize=3.0in\epsfbox{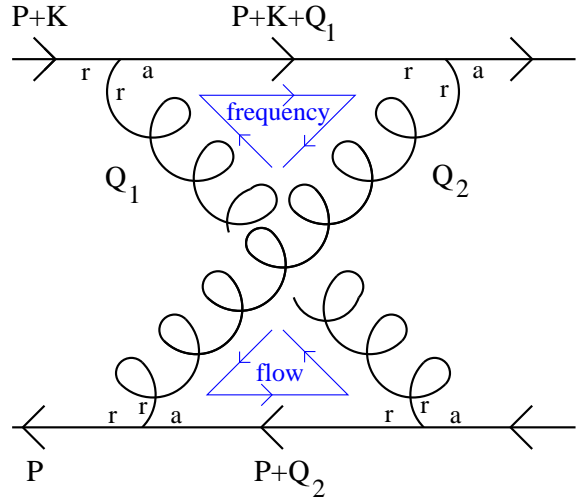}}
\vspace{0.1in}
\caption{\label{fig:cross} An individual loop in a cross-rung ladder.
The light arrows inside the crossed loop
illustrate the flow of frequency around the loop;
the fact that frequency flows from $a$ to $r$ on both scalar
propagators implies the absence of any ``pinching-pole'' enhancement.
}
\end{figure}

Next, consider the case where two gluonic lines cross,
as illustrated in Fig.~\ref{fig:cross}.
Define $\omega$ to be the {\em difference} between the frequencies on
the $Q_1$ and $Q_2$ gluonic lines, $\omega = q_1^0-q_2^0$, and
$\bar q^0$ as the corresponding average,
$\bar q^0 = \half q_1^0 + \half q_2^0$,
so that
$
dq_1^0 \> dq_2^0 = d\bar q^0 \> d\omega
$.
The $a,r$ index assignments are already forced on us by the need to have
pinching poles in the $P$ and $P{+}K$ scalar lines leading into
this part of the diagram, plus $rr$ assignments on the gluon rungs
in order to obtain Bose enhancement factors on the gluon lines.
The relevant part of the frequency integration thus becomes
\begin{eqnarray}
&& \int d\omega \> G_{ar}(K+P+Q_1) \, G_{ra}(P+Q_2)
=
\int d\omega \> G_\adv(K+P+Q_1) \, G_\ret(P+Q_2)
\\
    & = & \int d\omega \>
    \frac{1}{
    \Big[
	\left( k^0 + p^0 + \bar q^0
	+ \half \,\omega - \frac i2 \, \Gamma
	\right)^2
	- E_{\k{+}\p{+}\q_1}^2
    \Big]
    \Big[
	\left(-p^0 - \bar q^0
	+ \half \,\omega -\frac i2 \, \Gamma
	\right)^2
	- E_{\p{+}\q_2}^2
    \Big]
    } \, . \nonumber
\end{eqnarray}
We have written the denominator so that $\omega$ enters with
positive sign in each term.
Because $\omega$ is accompanied by $-i\Gamma$ in both terms,
the pole positions in the $\omega$ plane are all on the
same side of the real axis, and the contour can be deformed to
avoid all poles by a distance parametrically large compared to $\Gamma$.
Hence there is no pinch contribution,
and the crossed rung contribution is suppressed.
An easy way to see that this will happen is to note how the frequency
flows around the diagram, as illustrated in Fig.~\ref{fig:cross}.
The frequency flows through both quark propagators from left to right,
that is, from $a$ to $r$;
therefore (when one rewrites a retarded propagator with negative frequency
as an advanced propagator with positive frequency)
one gets a product of two advanced propagators,
rather than one retarded and one advanced.

Physically, the absence of crossed rungs can be understood most easily
by Fourier transforming from frequency to time.  The $a,r$ assignments
we have found necessary imply that the quark propagators,
from one current insertion to the other, go between vertices in
descending time order.
Hence a diagram like Fig.~\ref{fig:uncrossed} or 
Fig.~\ref{fig:cross}, read from left to right, goes
from later to earlier time.
Two neighboring, uncrossed gauge propagators 
are time ordered; both vertices of the
first propagator occur at a later time than either vertex of the second.
On the other hand, when two gauge boson lines cross, the top vertices are in
one time ordering and the bottom vertices are in another.  Since the time
scale for a soft scattering event is $\sim 1/gT$, while the typical time
between scatterings is $\sim 1/g^2 T$, it is natural that such
overlapping scattering events are suppressed.

\begin{figure}
\centerline{\epsfxsize=2.5in\epsfbox{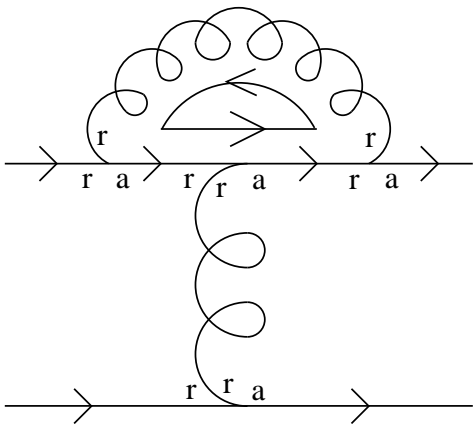}}
\vspace{0.1in}
\caption{A soft vertex correction, which is also sub-leading.\label{fig:cross2}}
\end{figure}

With this experience it is now easy to see why diagrams containing a soft
vertex correction like that in Fig.~\ref{fig:cross2} are also suppressed.
In terms of the added
frequency integration variable, both new quark propagators are traversed
from $a$ to $r$; the frequency integration will have two advanced
propagators and there will not be a pinching pole.

Hence, when all gauge boson lines carry soft $O(gT)$ momentum, ladder
diagrams of the form shown in Fig.~\ref{fig:ladder} contribute to the
leading order emission rate,
but crossed ladders of other forms or vertex corrections do not.

\subsection{Momentum exchanges softer than $gT$}
\label{sec:softer}

Our next task is to analyze the effects of gauge boson interactions
carrying momenta which are parametrically small compared to the
``soft'' $gT$ scale.
We will proceed in two stages,
first examining ``very-soft'' momentum scales which are intermediate between
$gT$ and $g^2 T$,
that is $|\q| = O(g^\nu T)$ for $1 < \nu < 2$.
We will argue that the previous analysis showing the dominance of
uncrossed ladder graphs continues to apply,
and then show that the net effect of these exchanges,
at leading order, is merely to cancel the dependence on
the contribution of momenta softer than $gT$ to the quark self-energy.
Then we will extend the discussion to ``ultra-soft'' $g^2 T$ scale
interactions, and argue that a similar cancellation continues to apply.

To begin, it will be useful to review properties of the
gauge boson propagator for momenta and frequencies
which are in the range $g^2 T \ll |\q| \ll gT$.
In this domain, the longitudinal part of the gauge field
propagator is screened at all frequencies,
so that for any spacelike $Q$,
\begin {equation}
    G^{\mu\nu}_{\ret,\rm L}(Q) \sim 1/\mD^2 = O(1/g^2 T^2) \,.
\end {equation}
However, the transverse self-energy \cite{Weldon}
has frequency dependence which will be essential.
For spacelike momenta with $|\q| \ll gT$,%
\footnote
    {%
    For timelike momenta, there is a value of $q^0$ where
    the (real part of the) inverse propagator has a zero,
    leading to a peak in $G^{\mu \nu}_{\ret}(Q)$;
    this is the plasmon.
    However, the plasmon pole occurs at 
    $q^0 \sim \mD = O(gT)$ even for $|\q|\ll gT$.
    For $|\q| \ll gT$, the spectral weight carried
    by the plasmon pole is small.
    Moreover,
    the frequency range where the scalar propagators are all near their
    pinching poles always involves spacelike four-momenta on all gluon lines,
    so the plasmon pole plays no role at leading order in $g$
    for either soft or very-soft gluons.
    }
\begin{equation}
    \Re \, \Pi^{\mu\nu}_{\ret,\rm T}(Q) \sim
    \left(\frac{q^0}{|\q|}\right)^2 \, (g T)^2 \, ,
\qquad
    \Im \, \Pi^{\mu\nu}_{\ret,\rm T}(Q) \sim \frac{q^0}{|\q|} \, (g T)^2 \, .
\end{equation}
If $q^0 \sim q$ then the transverse self-energy is
$O(g^2 T^2)$ and the transverse retarded propagator is $O(1/g^2T^2)$,
just like the longitudinal part.
But if $q^0 \sim |\q|^3/(gT)^2$,
then the self-energy is pure imaginary
[up to corrections suppressed by $\q^2/(g T)^2$]
and of order $\q^2$,
leading to a retarded propagator which is of order $1/\q^2$
--- larger than the longitudinal propagator by a factor of $(gT)^2/\q^2$.
This low (or ``quasi-static'') frequency regime dominates the spectral weight.
In the very soft region, the equal time, transverse gauge field
correlation function is given by its free theory value,
\begin {equation}
\label{eq:soft}
    i \int d^3x \> e^{-i \q \cdot \x} \,
    \langle A^\mu(\x,t{=}0) A^\nu(0,t{=}0) \rangle
    =
	\int \frac{dq^0}{2\pi} \> G^{\mu \nu}_{rr}(q^0,\q)
    \simeq
	\, i \: P_{\rm T}^{\mu \nu}(\hat\q) \, \frac{T}{\q^2} \, ,
\end{equation}	
with $P_{\rm T}^{\mu \nu}(\hat\q)$ denoting a transverse projector,
up to relative $O(\q^2 / g^2 T^2)$ corrections from the longitudinal
contribution and $O(g^2 T/|\q|)$ corrections from loops.%
\footnote
    {%
    The absence of a mass term in the transverse part is,
    of course, a consequence of gauge invariance.
    The form of \Eq{eq:soft} also has an interpretation in terms of
    dimensional reduction \pcite{FKRS}: the static limit of
    the transverse gauge boson correlator is given by
    ``dimensionally reduced'' 3-D Yang-Mills theory.
    }
The loop corrections become $O(1)$ at the scale $g^2 T$,
reflecting the fact that physics at this scale is non-perturbative.

When $|\q| = O(g^\nu T)$, with $\nu > 1$,
the $d^3\q$ phase space will contribute a factor of $(g^\nu T)^3$,
which is now parametrically smaller than the previous $(gT)^3$
soft phase space.
We will see, momentarily, that only the low frequency, transverse part of
the gauge boson propagator is large enough to produce a contribution in this
momentum region which will affect the leading order emission rate.
The longitudinal part of the propagator may be completely neglected.

In the previous section, we integrated over loop momenta
by doing the frequency integration first,
and argued that the only important frequency dependence
[in the $O(g^2 T)$ interval which dominated the integral]
was due to the scalar propagators.
But as just noted, for very-soft momenta the gauge field propagator has
significant frequency dependence on a scale of $|\q|^3/(gT)^2$,
which is of the order (or smaller) than $g^2 T$ for $\nu \ge 4/3$.
Hence, we must modify the previous pinching-pole analysis to handle
this regime where the gauge boson propagator depends strongly on $q^0$.
To do so, first recall that the poles which nearly pinch in the
scalar propagators $G_\adv(K{+}P{+}Q)$ and $G_\ret(P{+}Q)$
[{\em cf.} Fig.~\ref{fig:uncrossed}]
lie at $p^0{+}q^0 = \ppar{+}\qpar$, up to $O(g^2T)$ corrections.
Instead of using $\qpar$ and $q^0$ as independent variables in the
loop integration, we will use $q^- \equiv q^0{-}\qpar$ and $q^0$.
The essential point is that the scalar propagators depend only
weakly on $q^0$ if $q^-$ is held fixed.%
\footnote
    {%
    For fixed $q^-$,
    a variation in $q^0$ by $\delta q^0$
    changes the retarded propagator $G_\ret(P{+}Q)$ by an amount
    \begin {eqnarray}
	\delta G_\ret (P{+}Q)
	&=&
	2\delta q^0 \> {\partial \over \partial q^+} \> G_\ret(P{+}Q) 
    \nonumber\\ &=&
	2\delta q^0 \> [G_\ret(P{+}Q)]^2 \> {\partial \over \partial q^+}
	\left[
	    (p^-{+}q^-)(p^+{+}q^+) - (\p_\perp {+} \q_\perp)^2
	    - \Sigma_\ret(P{+}Q)
	\right]
    \nonumber\\ &=&
	2\delta q^0 \> [G_\ret(P{+}Q)]^2
	\left[
	    (p^-{+}q^-) - (\partial/\partial q^+) \, \Sigma_\ret(P{+}Q)
	\right] ,
    \label {eq:Gvar}
    \end {eqnarray}
    where
    $q^+ \equiv q^0 + \qpar$, and
    $
	\partial/\partial q^+ \equiv
	\half(\partial/\partial q^0 + \partial/\partial \qpar)
    $.
    The momentum derivative of the self-energy is order $g^2 T$
    provided $\p = O(T)$.
    And $(p^-{+}q^-) \, G_\ret(P{+}Q)$ is order $1/T$ both
    within the pinching-pole region
    [where $(p^-{+}q^-)$ is $O(g^2 T)$ and $G_\ret(P{+}Q)$ is $O(1/g^2 T^2)$],
    and outside this region.
    Hence, for fixed $q^-$, the relative variation in the retarded propagator
    $
	{\delta G_\ret / G_\ret} = O(\delta q^0 /T)
    $.
    Exactly the same estimate holds for the advanced propagator
    $G_\adv(K{+}P{+}Q)$.
    Provided $q^0$ varies by $O(gT)$ or less,
    this implies that if we treat the scalar propagators
    as depending on $q^0$ and $\qpar$ only through the combination $q^-$,
    then we will be making at most an $O(g)$ error
    which is irrelevant to our leading-order analysis.
    }
Consequently, the integral over $q^0$, at fixed $q^-$, will be
controlled by the frequency dependence of the gauge field
propagator $G^{\mu\nu}_{rr}(Q)$, not the scalar propagators,
and receive its dominant contribution from the low frequency regime
where $q^0 = O(|\q|^3/g^2T^2)$.
The nearly pinching poles in the scalar propagators will then
appear in the $q^-$ integral.

The resulting parametric estimate for the relative size
of a very-soft ($|\q| \sim g^\nu T$) gauge boson exchange is
\begin{eqnarray}
    &&
    g^2 T^2 \int_{|\q| \sim g^\nu T} {d^4Q \over (2\pi)^4} \;
    G^{\mu \nu}_{rr}(Q) \, G_\ret(P{+}Q) \, G_\adv(K{+}P{+}Q)
\nonumber\\ &\sim&
    g^2 T^2
    \int_{q_\perp \sim g^\nu T} {d^2 q_\perp \over (2\pi)^2}
    \left[ \int {dq^- \over 2\pi} \>
	G_\adv(K{+}P{+}Q) \, G_\ret(P{+}Q) \Bigr|_{q^0=0}
    \right]
    \left[ \int {dq^0 \over 2\pi} \>
	G^{\mu \nu}_{rr}(Q) \Bigr|_{q^-=-p^-}
    \right]
\nonumber \\
    &\sim&
	\left. g^2T^2 \times q^2 \times (1/g^2 T^3) \times (T/q^2)
	\right|_{q \sim g^\nu T}
\nonumber \\
    &\sim&
	1 \, .
\label {eq:very-soft}
\end{eqnarray}
The $q^0$ integral, more explicitly, is
$
    \int (dq^0/2\pi) \> G^{\mu\nu}_{rr}(q^0,\qpar{=}q^0{-}q^-,\q_\perp)
$
and coincides, to leading order,
with the static result (\ref{eq:soft}) for $|\q| \ll gT$
since the dominant contribution arises from $O(|\q|^3/g^2T^2)$
frequencies which are small compared to $|\q|$.
Hence the difference between holding $\qpar$ or $q^-$ fixed
when integrating over $q^0$ is subdominant.
The $q^-$ integral over the product of scalar propagators
gives an $O(1/g^2T^3)$ result whose explicit form is
just the appropriate adaptation of \Eq{eq:pinch},
namely
\begin{equation}
\label{eq:pinch2}
    \int \frac{dq^-}{2\pi} \>
    G_{\adv}(K{+}P{+}Q) \, G_{\ret}(P{+}Q) \biggr|_{q^0=0}
    \simeq
    \frac{1}
    {
    4 p^0 \, (p^0{+}k) \, (\Gamma + i \, \delta E)
    }
\end{equation}
[with $\delta E$ given in \Eq{eq:dEdef2}],
provided $|\q| \ll |\p|$ and $p^0 \simeq \ppar$.

The estimate (\ref{eq:very-soft}) implies that a very-soft
gauge boson exchange with momentum $g^2 T \ll |\q| \ll gT$
can be just as important as a soft $O(gT)$ exchange.
Of course, if one very-soft gauge boson exchange can make an $O(1)$
contribution to the emission rate, so will uncrossed ladders containing
any number of such exchanges.
Crossed rungs remain negligible in this very-soft regime
for the same reason as before;
if both propagators in the $q^-$ integration are
retarded or both are advanced, there is no pinching contribution.

It is important to note that the very soft momentum region also
makes an $O(1)$ contribution to the value of the on-shell scalar self-energy,
which controls the size of the pinching pole contribution (\ref {eq:pinch2}).
It therefore appears that momenta softer than $gT$ are important at
two points in the calculation: self-energies and uncrossed ladders.
However, we now show that these two contributions from momenta $q \ll gT$
actually cancel to leading order in $q^2/g^2 T^2$.

\begin{figure}[t]
\centerline{\epsfxsize=6.5in\epsfbox{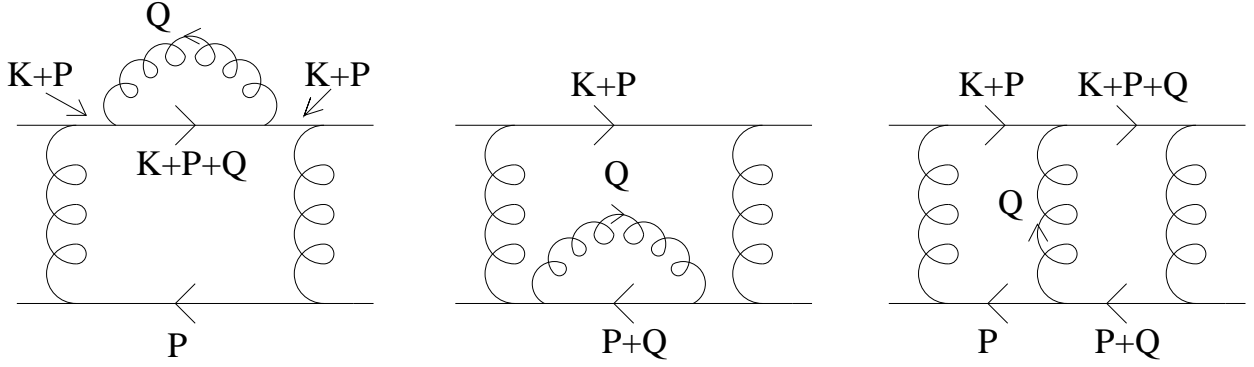}}
\vspace{0.1in}
\caption{\label{fig:cancel}Three contributions involving the addition
of a very-soft gauge boson, $|\q| \ll gT$, which cancel to leading order.
}
\end{figure}

To understand this cancellation, consider the sum of the three
sub-diagrams shown in Fig.~\ref{fig:cancel}, representing the
contribution of a particular momentum $\q$ to the scalar self-energies
of a pinching pair of propagators and to a possible cross-rung between them.
At this point, one should imagine having introduced a separation scale $\mu$
intermediate between $g^2 T$ and $gT$, and using propagators in which
self-energy contributions from all scales larger than $\mu$ have been
resummed, but where self-energy corrections from scales softer than $\mu$
are treated as explicit loop corrections.

To start, note that the momentum flowing along either leg
on the right side of the fragments shown in Fig.~\ref{fig:cancel}
must differ by $\pm Q$ for the case of very-soft gluon exchange
as compared to either of the self-energy insertions.
But this is a small correction to the rest of the diagram;
adding a very-soft momentum $\q$ to $\p$ changes the parallel (to $\k$)
component of $\p$ negligibly since $\ppar = O(T)$,
and changes the orthogonal component $\p_\perp \sim gT$
by a fractional amount which is at most $O(|\q|/gT)$.
In fact, once one averages over the direction of $\q_\perp$,
the relative size of surviving corrections is only
$O(\q_\perp^2 / \p_\perp^2) \sim q^2 / g^2 T^2$.
Hence we need only check whether the
contributions of the sub-diagrams pictured cancel.
These contributions, including the $p^0$ and $q^0$ frequency integrals
but neglecting common overall factors, are
\begin{eqnarray}
\label{eq:mess1}
    &&
	g^2
	\int \frac{dp^0}{2\pi}
	\int \frac{dq^0}{2\pi} \:
	G_{rr}^{\mu \nu}(Q)
\nonumber \\ && \quad {} \times
	\Bigg\{
	    (2K{+}2P{+}Q)_\mu (2K{+}2P{+}Q)_\nu \,
	    [G_\adv(K{+}P)]^2 \, G_\adv(K{+}P{+}Q) \, G_\ret(P)
\nonumber \\ && \quad \quad {} +
	    (2P{+}Q)_\mu(2P{+}Q)_\nu \,
	    G_\adv(K{+}P) \, [G_{\ret}(P)]^2 \, G_\ret(P{+}Q)
\nonumber \\ && \quad \quad {} +
	    (2K{+}2P{+}Q)_\mu (2P{+}Q)_\nu  \,
	    G_\adv(K{+}P) \, G_\adv(K{+}P{+}Q) \, G_\ret(P) \, G_\ret(P{+}Q)
	\Bigg\} \, .
\end{eqnarray}
At leading order, all the terms of form $(2K{+}2P)_\mu$ are parallel
with $K$ and may be replaced with their magnitudes times $\hat{K}^\mu$,
where $\hat{K}\equiv(1,\hat\k)$.
Further, $|\q+\k+\p| \simeq |\k + \p| \simeq |k+\ppar|$.
The scalar propagators depend weakly on $q^0$ for fixed $q^-$,
so the $q^0$ integral may be performed as discussed above.
Next, we may approximate each scalar propagator by just
the relevant nearly-pinching single pole.
For convenience, we will also define a shifted frequency variable,
$p'^0 \equiv p^0 - E_\p \, {\rm sign}(\ppar)$,
and then drop the prime so that
\begin{eqnarray}
    G_{\ret}(P) &=&
    \frac{1}{E_\p^2 - [p^0 +i\Gamma/2]^2}
    \Rightarrow
	-\frac{1}{2 E_\p \, \sgn(\ppar)}\; \frac{1}{p^0+i\Gamma/2} \, ,
\\
    G_{\adv}(K{+}P) &=& \frac{1}{E_{\p{+}\k}^2 - [p^0 {+} k^0 -i\Gamma/2]^2}
    \Rightarrow
	-\frac{1}{2E_{\p{+}\k} \, \sgn(\ppar{+}k)} \;
	\frac{1}{p^0 + \delta E - i\Gamma/2} \, ,
\end{eqnarray}
and likewise for all the other propagators.
The key observation is that the $\q_\perp$ dependence of the
difference between the pole positions of $G_\ret(P)$ and $G_\ret(P{+}Q)$,
or between $G_\adv(K{+}P)$ and $G_\adv(K{+}P{+}Q)$,
are negligible provided $\q \ll gT$.
In other words,
\begin{eqnarray}
\label{eq:dEdq}
    \delta E &\equiv&
    \sgn(\ppar) \, E_\p
    - \sgn (\ppar{+}k) E_{\p{+}\k} + k^0
\nonumber\\ &\simeq&
    \sgn(\ppar) \, E_{\p}
    - \sgn(\ppar{+}\qpar{+}k) \, E_{\p{+}\k{+}\q} + k^0 + \qpar \, ,
\nonumber\\ &\simeq&
    \sgn(\ppar{+}\qpar) \, E_{\p{+}\q}
    - \sgn(\ppar{+}k) \, E_{\p{+}\k} + k^0  - \qpar \, ,
\nonumber\\ &\simeq&
    \sgn(\ppar{+}\qpar) \, E_{\p{+}\q}
    - \sgn(\ppar{+}\qpar{+}k) \, E_{\p{+}\k{+}\q} + k^0 \, .
\end{eqnarray}
This amounts to neglecting the contribution of 
$\q_\perp^2 + 2\p_\perp \cdot \q_\perp$ to $(\p_\perp{+}\q_{\perp})^2$.
The corrections, relative to the size of $\delta E$, are of order
$\q_\perp^2/g^2 T^2$ after averaging over the direction of $\q_\perp$,
since $\p_\perp^2 \sim g^2 T^2$ and $\p_\perp \cdot \q_\perp$ vanishes upon
angular integration.

Under these approximations,
and recalling that $q^0 \ll q^- \simeq \qpar$ for very-soft $\q$,
expression (\ref{eq:mess1}) becomes
\begin{eqnarray}
\label{eq:mess2}
    &&
    g^2
    \int \frac{dp^0}{2\pi} \>
    \PT^{\mu \nu}(\hat\q) \, \frac{T}{\q^2} \:
    \frac{\hat{K}_\mu \hat{K}_\nu}{4\ppar (k{+}\ppar)}
    \left[
	\frac{1}{( p^0 {+} i\Gamma/2)(p^0{-}i\Gamma/2{+}\delta E)}
    \right]
\nonumber \\ && \qquad {}
    \times
    \left\{
	\frac{1}{(p^0{-}i\Gamma/2{+}\delta E) \,
	(p^0{-}\qpar{-}i\Gamma/2{+}\delta E)}
    \right.
	+ \frac{1}{(p^0{+}i\Gamma/2) \,
	(p^0{-}\qpar{+}i\Gamma/2)}
\nonumber \\ && \qquad \quad
    \left.
	+ \frac{1}{(p^0{-}\qpar{-}i\Gamma/2{+}\delta E) \,
	(p^0{-}\qpar{+}i\Gamma/2) }
    \right\} .
\end{eqnarray}
We now perform the $p^0$ integral.  The first two terms have one pole
on one side of the contour and three on the other; the last term has two
poles on each side, so it gives one contribution for each residue.
The result is proportional to
\begin{eqnarray}
    &&
    \frac{1}{i\Gamma {-} \delta E}
    \left\{
	\frac{1}{(i\Gamma{-}\delta E + \qpar)}
	\left[
	    \frac{1}{(i\Gamma{-} \delta E)} + \frac{1}{\qpar}
	\right]
	+
	\frac{1}{(i\Gamma{-}\delta E - \qpar)}
	\left[
	    \frac{1}{(i\Gamma{-} \delta E)} - \frac{1}{\qpar}
	\right]
    \right\}
	= 0 \, .
\label{eq:cancel}
\end{eqnarray}
Corrections to this cancellation arise from the 
corrections to the approximate equalities (\ref{eq:dEdq}),
as well as corrections from neglecting $Q$ (relative to $P$)
in other parts of the diagram.
All such corrections are suppressed,
after averaging over the direction of $\q_\perp$, by
at least $O[\q^2 / (gT)^2]$.
In particular this cancellation is {\em not} sensitive to corrections to
the static correlator (\ref{eq:soft}),
and so is unaffected by the loss in perturbative calculability
of this correlator when $|\q| = O(g^2 T)$.

To be complete,
we should also check that the three diagrams of Fig.~\ref{fig:cancel}
have the same group theoretic factor in a non-Abelian theory.
This turns out to be trivial.
In any diagram where the gluonic lines never cross
(in the sense we have used above),
so called ``rainbow topologies,'' the non-Abelian group factor is%
\footnote
    {%
    Note that this would be false,
    and the cancellation (\ref {eq:cancel}) would not occur,
    if we were considering color conductivity \pcite{Bodeker,ASY}.
    }
$(\cf)^{\# {\rm gluons}}$.

The physical interpretation of the cancellation (\ref {eq:cancel})
is as follows.  
There is an $O(1)$
chance that during the order $1/g^2 T$ time required for the photon emission
process to complete, there will be a scattering with exchange of $q \ll
gT$ momentum.
However, such a scattering disturbs the quark (or scalar) too little to
disrupt the emission process.
The self-energy correction accounts for the removal of the quark (or scalar)
from the initial momentum state, and the ladder rung accounts for
its contribution in the new scattered state.
This interpretation will become more manifest when we resum ladder diagrams
in the next section.
In a relativistic context, this cancellation 
is the same as that discussed
by Lebedev and Smilga \cite{LebedevSmilga} in the context of electrical
conductivity
(see also Ref.~\cite{Bodeker2}), but the basic observation goes back
decades.%
\footnote
   {%
    A single small-angle scattering does not substantially affect
    transport processes such as electrical conductivity.  In the
    non-relativistic context, this insensitivity to
    small angle scattering appears as a suppression factor of 
    $(1 - \cos \theta)$ in what is sometimes called the
    ``transport cross-section'' \cite{textbook1}.  This factor
    is relatively easily derived from the Boltzmann equation.  
    Diagrammatically, it emerges as a cancellation, when
    summing ladder diagrams, between rung and self-energy insertions.
    An early discussion may be found in Ref.~\cite{Edwards};
    see also Sec.~39.2 of Ref.~\cite{Abrikosov}.
    }
However, these works address a different kinematic regime than
the one relevant here (they consider extremely soft $k$, whereas we
have hard but lightlike $k$), and the range of exchange momentum for
which the cancellation occurs in the current context is not the same as
in these references.  In particular, in our application
an exchange momentum $|\q| \sim gT$ is enough to disturb the collinearity
condition which must be satisfied for the poles to pinch.  This is not
the case for softer $k$.

Having found that the leading order photon emission rate is insensitive to
the effects of very-soft ($g^2 T \ll |\q| \ll gT$) gauge bosons,
one should naturally expect the emission rate to be equally
insensitive to the ultra-soft $g^2 T$ scale.
(Otherwise, one would have a peculiar situation in which a physical
quantity, at leading order, depends on the dynamics of two different
momentum scales, $gT$ and $g^2 T$, but not on intervening scales.)
Certainly, the physical picture just sketched does not suddenly
change when $\q$ reaches the $g^2 T$ scale.

\begin{figure}
\centerline{\epsfxsize=4.5in\epsfbox{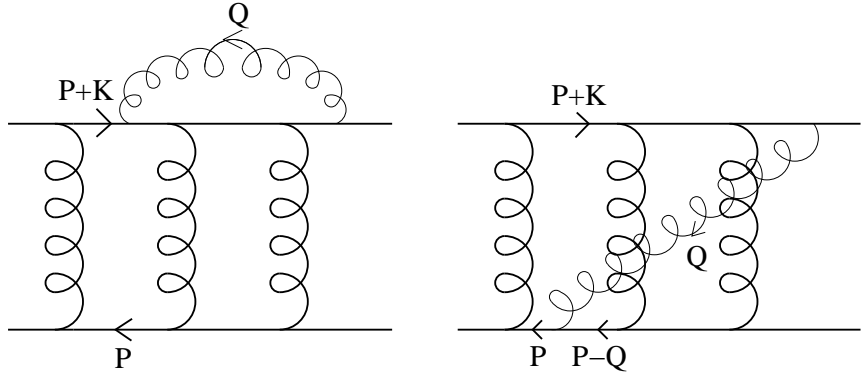}}
\vspace{0.1in}
\caption{\label{fig:g2T} Two ways of adding a $g^2 T$ line, denoted by
the thin gluon line with momentum $Q$, which will cancel at leading order.}
\end{figure}

However, some readers may want a more deductive demonstration
that this insensitivity remains valid at the $g^2 T$ scale.
To this end, first observe that
nothing changes in the analysis of the diagrams of Fig.~\ref{fig:cancel}
if the momentum $\q$ is in the ultrasoft $O(g^2T)$ regime.
The cancellation among these diagrams remains applicable at the
non-perturbative $g^2 T$ scale.
However, in this regime, the previous arguments which ruled out
crossed diagrams and vertex corrections no longer apply;
one should verify that a similar cancellation also occurs between
diagrams of the form shown in Fig.~\ref{fig:g2T}.
The demonstration is very similar to that for uncrossed graphs.

It is sufficient to show that when we sum over the two possibilities
of attaching the left vertex of the $g^2T$ gluon line to either the
upper or lower quark lines, there is an $O(g)$ cancellation.
The same argument will apply, independently, for the attachment point
of the right vertex, leading to an overall cancellation by $O(g^2)$.
The cancellation we will find also applies if the other end of the $g^2T$
line attaches to another ultrasoft gluon line,
so it will rule out more complicated ultrasoft corrections as well. 

First note that the addition of the ultrasoft $\q \sim g^2 T$ line does
not modify, at leading order, any of the integrations over loops which
the ultrasoft propagator skips over.  This is because the transverse
momentum it carries is smaller by $g$ than the typical transverse
momentum already in the loop, and the longitudinal momentum is smaller
by $g^2$.  Therefore we need only focus on the pair of pinching poles which
are augmented by one more propagator when (one end of) the ultrasoft line
attaches.
The sum of the two contributions at fixed $\q$, after performing the
$q^0$ integration
and suppressing factors of $g$ and group factors, is
\begin{eqnarray}
&& 
    \PT^{\mu \nu}(\q) \, \frac {T}{\q^2}
    \int \frac{dp^0}{2\pi} \>
    \bigg[\,
	2(P{+}K)_\mu \, G_\adv(P{+}K) \, G_\adv(P{+}K{+}Q) \, G_\ret(P)
\nonumber \\ && \hspace{1.3in} {}
	+ 2 P_\mu \, G_\adv(P{+}K) \, G_\ret(P{-}Q) \, G_\ret(P)
    \,\bigg]
\nonumber \\ &\simeq&
 -  \frac{\hat{K}_\mu \PT^{\mu \nu}(\q)}{4\ppar(k{+}\ppar)} \,
    \frac {T}{\q^2}
    \int \frac{dp^0}{2\pi}
	\left[
	    \frac{1}{(p^0{-}i\Gamma/2{+}\delta E) \,
	    (p^0{-}i\Gamma/2{+}\delta E{+}\qpar) \, (p^0{+}i\Gamma/2)}
	\right.
\nonumber \\ && \hspace{1.4in}
	\left. {}
	    + \frac{1}{(p^0{-}i\Gamma/2{+}\delta E) \,
	    (p^0{+}i\Gamma/2{-}\qpar) \,(p^0{+}i\Gamma/2)} \,
	\right]
\nonumber \\ &=&
    \frac{\hat{K}_\mu\PT^{\mu \nu}(\q)}{4\ppar(k{+}\ppar)} \,
    \frac {T}{\q^2}
	\left[ \frac{1}{(i\Gamma{-}\delta E)(i\Gamma{-}\delta E
	{-}\qpar)} - \frac{1}{(i\Gamma{-}\delta E)
	(i\Gamma{-}\delta E{-}\qpar)} \right] ,
\end{eqnarray}
and again the two terms cancel.
(In the first step, we shifted $p^0$ by $E_\p$ and retained
only the pinching pole terms, exactly as before.)
And just like before, one can verify that non-Abelian
group factors do not affect this cancellation
(because the currents at either end of the full diagram
are gauge invariant).
%

\begin{figure}
\centerline{\epsfxsize=2.4in\epsfbox{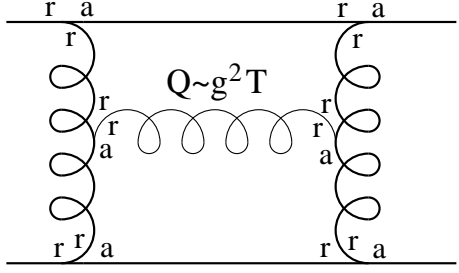}}
\vspace{0.1in}
\caption{A diagram fragment in which an ultra-soft gluon
line connects two soft gluon rungs.
Such diagrams are suppressed by $O(g^2)$.
\label{fig:g2T2}}
\end{figure}

It is elementary to see that there are also no leading order
contributions when an ultrasoft line attaches to one of the soft
exchange gluons.  This is because the vertex involves a
derivative coupling and only an order $gT$ momentum runs
through the soft gauge boson line.
To be definite, consider an ultrasoft gauge boson connecting two soft 
gauge boson lines, as shown in Fig.~\ref{fig:g2T2}.  
No new pinching poles are added.  There is an
explicit $g^2$ from the new vertices,
a $g^2$ from the derivatives at the vertices,
a $g^6$ from the ultrasoft phase space,
a $1/g^4$ from the frequency integration over the ultrasoft $rr$ propagator,
and $1/g^4$ from two new $ra$ soft gauge boson propagators.
Putting it together, there is a $g^2$ suppression.
Physically,
this reflects the fact that the soft exchange gluons only
propagate a distance $1/gT$, much shorter than their mean free path to
interact with the ultrasoft field which is of order $1/g^2 T$.

Therefore, although scales softer than $gT$ appear to contribute
if one examines individual diagrams,
when appropriate combinations of diagrams are summed, contributions
from these scales cancel.
There is therefore no danger in temporarily
imposing an infrared regulator, whose form we need not specify,
which eliminates the sensitivity of the decay width $\Gamma_{\p}$ to
ultrasoft physics
and allows the manipulations of the next section.
After summing the required diagrams, the infrared regulator may
be removed without re-introducing any poor behavior
({\em i.e.}, infrared sensitivity).

\section{Summing ladder diagrams}
\label{sec:resum}

\begin{figure}[t]
\centerline{\epsfxsize=6.3in\epsfbox{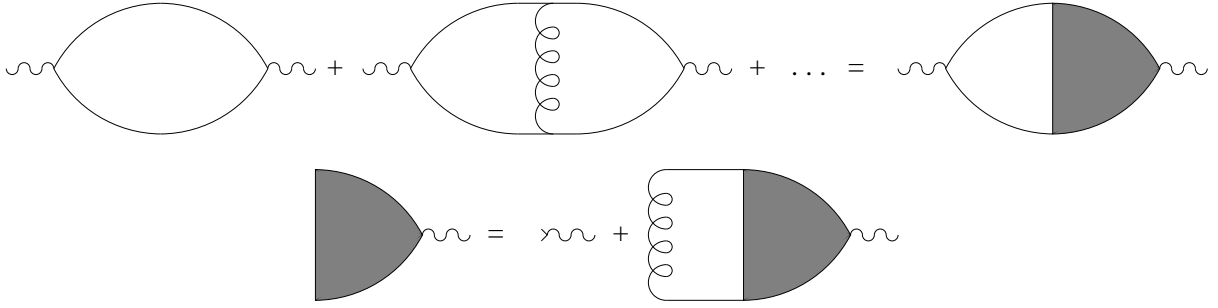}}
\vspace{0.1in}
\caption{\label{fig:schwinger} Summation of the geometric series
of ladder diagrams.}
\end{figure}

\begin{figure}[t]
\centerline{\epsfxsize=3.6in\epsfbox{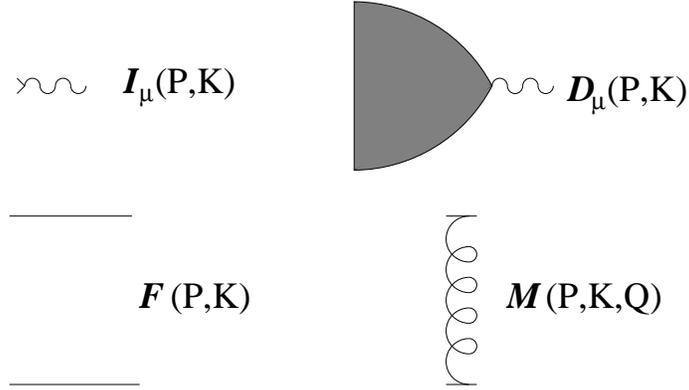}}
\vspace{0.1in}
\caption{\label{fig:pieces}
Correspondence between graphical elements and the symbols
appearing in the integral equation which represents the sum of all
ladder graphs.
}
\end{figure}

Summing the contributions of all ladder diagrams requires a procedure
analogous to the formulation of a Bethe-Salpeter equation;
one cuts off the photon vertex from one end of every diagram,
recognizes that the resulting sum is just a geometric series,
and writes down the linear integral equation whose iteration
generates this series.
This is illustrated in Figs.~\ref{fig:schwinger} and \ref{fig:pieces}.

To carry this out explicitly, we first introduce some notation,
chosen to mimic the treatment of Jeon \cite {Jeon} as much as possible.
We define a bare photon vertex ${\cal I}^\mu(P;K)$,
as well as its transverse projection ${\cal I}^\mu_\perp(P;K)$,
\begin{equation}
    {\cal I}^\mu(P;K) \equiv (2P{+}K)^\mu \,, \qquad
    {\cal I}^\mu_\perp(P;K) \equiv (2P{+}K)^\mu_\perp = 2P^\mu_\perp \, ,
\label {eq:Idef}
\end{equation}
which represent the vertex factors associated with the attachment
of an external photon.
The full ${\cal I}^\mu$ will be used to find the complete current-current
correlator,
while ${\cal I}^\mu_\perp$ will generate the transverse projected
correlator, which turns out to provide the most economical form
for extracting the photon production rate.
We will want both in what follows,
in order to check explicitly that our leading-order result
for the current-current correlator is transverse.

We introduce ${\cal F}(P;K)$ to denote the pinching-pole part of
a pair of scalar propagators, 
with $r,a$ indices appropriate for evaluating $G_{aarr}$;
\begin{eqnarray}
	{\cal F}(P;K)
    &=&
	\left.
	    (-i)^2 \, G_\adv(P{+}K) \, G_\ret(P)
	\right|_{\rm pinching-pole} \, ,
\nonumber \\
    &\equiv&
	\frac{-1}{4\ppar \, (\ppar{+}k)} \:
	\frac{1}{\Gamma+i\delta E} \:
	4 \pi \,
	\delta\Big[
	    2p^0+k^0 -E_{\p}\, \sgn(\ppar) - E_{\p{+}\k}\, \sgn(\ppar{+}k)
	\Big] \, ,
\label {eq:Fdef}
\end{eqnarray}
where the second expression is the pinching pole approximation,
and $\Gamma$ is shorthand for $\half (\Gamma_\p {+} \Gamma_{\p{+}\k})$.
The delta function represents the product of nearly pinching single poles,
as in \Eq{eq:pinch},
and so should really be smeared out over the $O(g^2 T)$ width
of the pinching-pole region.
But since the exchange momenta are $O(gT)$,
approximating the product of (off-axis) poles
by the indicated delta function is sufficient.
We have also, once again, used the fact that $\ppar$ will be $O(T)$
while $\p_\perp$ is $O(gT)$ to write $\ppar(\ppar{+}k)$
in the prefactor instead of $E_\p E_{\p+\k}$ (times an appropriate $\pm$ sign).

Next, we introduce a linear operator $\cal M$ which will represent
the addition of a gauge boson cross-rung with $rr$ indices.
The kernel of this operator is
\begin{equation}
    {\cal M}(P,Q;K) \equiv 
    i \cf \, g^2
    \, 2P_\mu \> 2(K{+}P)_\nu \, G^{\mu \nu}_{rr}(Q) \, .
\end{equation}
The overall factor of $i$ reflects the
$i$'s associated with each vertex (\ref {eq:vertex})
together with one $-i$ for the propagator to compensate
for the $i$ included in the definition (\ref{eq:Grr-def}) of $G_{rr}$.
We have already made a $Q \ll P,K$ approximation in writing $2P_\mu$ rather
than $(2P{+}Q)_\mu$, etc.
We may also use the lightlike nature of $K$ and
(near) collinearity of $P$ and $K$ to further simplify the kernel to the form
\begin{equation}
    {\cal M}(P,Q;K) = i g^2 \cf \, 4\ppar(\ppar{+}k)  \,
    \hat{K}_\mu \hat{K}_\nu \, G^{\mu \nu}_{rr}(Q) \, ,
\label {eq:Mdef}
\end{equation}
with $\hat K^\mu = (1, \hat \k)$ as before.
Note that the group factors are trivial since
none of the gluon lines cross;
there is just a factor of $g^2 \cf$ for each gauge boson line.

Finally, we introduce an effective vertex ${\cal D}^\mu(P;K)$,
which represents the sum of all numbers of cross-rungs and pinching-pole
propagators placed in front of the photon vertex ${\cal I}^\mu$,
\begin {equation}
    {\cal D}^\mu \equiv 
    {\cal I}^\mu
    + {\cal MF I}^\mu
    + {\cal MFMF I}^\mu
    + {\cal MFMFMF I}^\mu
    + \cdots \,.
\end {equation}
More explicitly,
\begin{eqnarray}
    {\cal D}^\mu(P;K)
    &=&
    {\cal I}^\mu (P;K) 
\nonumber \\ && {}
    + \int_{Q_1} {\cal M}(P,Q_1;K)
	\, {\cal F}(P{+}Q_1;K)
	\, {\cal I}^\mu(P{+}Q_1;K)
\nonumber \\ && {}
    + \int_{Q_1} {\cal M}(P,Q_1;K) \, {\cal F}(P{+}Q_1;K) 
      \int_{Q_2} {\cal M}(P{+}Q_1,Q_2;K) \, {\cal F}(P{+}Q_1{+}Q_2;K)
\nonumber \\ && \hspace{3.0in} {} \times 
	{\cal I}^\mu(P{+}Q_1{+}Q_2;K)
\nonumber \\ && {}
	+ \, \cdots \,.
\end{eqnarray}
As illustrated in Fig.~\ref{fig:schwinger},
this is alternately expressed as the linear integral equation
\begin{equation}
\label{eq:integral}
    {\cal D}^\mu(P;K) = {\cal I}^\mu(P;K) + \int_Q {\cal M}(P,Q;K)
	\, {\cal F}(P{+}Q;K) \, {\cal D}^\mu(P{+}Q;K) \, .
\end{equation}
As also illustrated in Fig.~\ref{fig:schwinger},
the resulting LPM contributions to the full current-current correlator are
proportional to the real part of
an ``inner product'' of ${\cal IF}$ with ${\cal D}$,
\begin{equation}
\label{eq:jj}
    W^{\mu \nu}_\LPM(K)
    = -\int \frac{d^4 P}{(2\pi)^4} \>
    2 n(p^0{+}k^0) \, [1{+}n(p^0)]
    \: {\cal I}^\mu (P;K) \: \Re \left[ \strut
	{\cal F}(P;K) {\cal D}^\nu(P;K) \right] \, .
\end{equation}
The overall minus sign is present because the population functions,
as given by \Eq{eq:alphabeta}, are $n(p^0{+}k^0)n(-p^0)$,
and $n(-p^0)=-[1{+}n(p^0)]$.
The fact that only the real part enters follows from the discussion after
\Eq{eq:alphabeta}.  Although it is not obvious from its structure,
\Eq{eq:jj} does yield a current-current correlator $W^{\mu\nu}(K)$
which is symmetric under interchange of $\mu$ and $\nu$.
This will be shown in Sec.~\ref{sec:transverse}.

As it stands, \Eq{eq:integral} is a four-dimensional integral equation
characterizing the $P$ dependence of ${\cal D}^\mu(P;K)$.
However, the delta-function in ${\cal F}(P;K)$ implies that
${\cal F}(P;K)\, {\cal D}^\mu(P;K)$ has support only for a single
frequency $p^0$ for a given spatial momenta $\p$ (and $K$).
Consequently, it is convenient to define 
\begin{equation}
    f^\mu(\p;\k) \equiv - 4\ppar \, (\ppar{+}k)
    \int \frac{dp^0}{2\pi} \: {\cal F}(P;K) \, {\cal D}^\mu(P;K) \, ,
\end{equation}
in terms of which the current-current correlator (\ref{eq:jj})
becomes
\begin{equation}
\label{eq:jj_final}
    W^{\mu\nu}_\LPM(K)
	= \int \frac{d^3 \p}{(2\pi)^3} \>
	\frac{2 n(p^0{+}k^0) \, [1{+}n(p^0)]}{4\ppar \, (k{+}\ppar)}
	\: \left( 2P^\mu {+}K^\mu \right)
	\Re \, f^\nu(\p;\k) \, .
\end{equation}
In this equation, and in the equations which follow, the time components
of 4-vectors are to be understood as dependent parameters, fixed by the
locations of pinching pole pairs; so in the above, 
$2p^0 {+} k^0 = E_{\p}\, \sgn(\ppar) + E_{\p{+}\k} \, \sgn(\ppar{+}k)$.

The integral equation (\ref{eq:integral}) may be reduced to
an equation just involving $f^\mu(\p;\k)$.
To do so,
multiply \Eq{eq:integral} by ${\cal F}(P;K)$,
integrate over $p^0$, and
then multiply both sides by $4\ppar(\ppar{+}k)(\Gamma{+}i\delta E)$.
This yields
\begin{eqnarray}
\label{eq:f1}
    2P^\mu {+} K^\mu
    &=&
    (i \delta E + \Gamma) \, f^\mu(\p;\k)
	- \int \frac{d^3 \q}{(2\pi)^3} \;
	{\cal C}(\q;\k) \, f^\mu(\p{+}\q_\perp;\k) \, ,
\end{eqnarray}
where
\begin {equation}
    {\cal C}(\q;\k)
    \equiv
    g^2 \cf \int \frac{dq^0}{2\pi} \> 2\pi \delta(q^0{-}\qpar)
    \left[ -i \, G^{\mu \nu}_{rr}(Q) \hat{K}_\mu \hat{K}_\nu \right]  .
\label {eq:Cdef}
\end {equation}
The replacement of $(\p{+}\q)$ with $(\p{+}\q_\perp)$ in the first
argument of $f^\mu$ inside the $d^3\q$ integral is justified at leading
order because $\qpar$ is negligible compared to $\ppar$,
and the variation of $f^\mu(\p;\k)$ with $\ppar$ occurs on a scale of $T$.
[In contrast, the $f^\mu(\p;\k)$ varies with $\p_\perp$ on the $gT$ scale.]
We have also made the approximation
$E_{\p{+}\q} {-} E_{\p} = \qpar \, \sgn(\ppar)$, valid at leading order, to
turn the delta function in ${\cal F}(P{+}Q,K)$ 
into $\delta(q^0{-}\qpar)$.

Next, we need to insert the actual form of the damping rate $\Gamma$.
It is related to the imaginary part of the on-shell retarded self-energy,
\begin{equation}
    p^0 \, \Gamma_\p = g^2 \cf \: \Im \int \frac{d^3 \q\, dq^0}{(2\pi)^4} \>
	4P_\mu P_\nu \left[-iG^{\mu \nu}_{rr}(Q)\right] G_\ret(P{+}Q) \, .
\end{equation}
Everything is real except the retarded propagator; its imaginary part is
half the quark spectral weight, 
\begin{equation}
    \Im \: G_\ret(P{+}Q)
    \simeq \pi \, \delta [(P{+}Q)^2]
    \simeq \frac{2 \pi \delta (q^0{-}\qpar)}{4 p^0} \, ,
\end{equation}
where we used the on-shell condition $p^0 = E_\p \sgn(\ppar)$
and $dE_{\p{+}\q}/d\q \simeq \hat\p \simeq \hat\k$,
dropped the irrelevant pole at $p^0 \simeq - E_\p \sgn(\ppar)$, 
and took the quasiparticle width to be small compared to $q$,
which is legitimate for $q \sim gT$.
(Remember that the $g^2T$ scale is irrelevant and we have implicitly
regulated it.)
Consequently,
\begin{equation}
\label{eq:Gamma}
\Gamma = g^2 \cf \int \frac{d^3 \q \, dq^0}{(2\pi)^4} \>
	2\pi \delta(q^0{-}\qpar) \left[ -i \, G^{\mu \nu}_{rr}(Q)
	\hat{K}_\mu \hat{K}_\nu \right] \, ,
\end{equation}
and therefore we can rewrite \Eq{eq:f1} as
\begin{eqnarray}
\label{eq:f2}
    2P^\mu{+}K^\mu
    &=&
    i \delta E \, f^\mu(\p;\k)
	+ \int \frac{d^3 \q}{(2\pi)^3} \;
	{\cal C}(\q;\k) \,
	\Big[ f^\mu(\p;\k) -  f^\mu(\p{+}\q_\perp;\k) \Big] \, .
\end{eqnarray}
This is structurally similar to a Boltzmann equation;
in real space the $i \delta E$ term can be viewed
as a gradient (flow) term,
and $\cal C(\q;\k)$ as the rate for scattering with momentum transfer $\q$.
The self-energy has turned into a loss term, accounting for the scattering
of excitations out of momentum $\p$, and the ladder exchanges have generated
a compensating gain term, accounting for scattering into momentum $\p$
from momentum $\p{+}\q$.  

It remains to present the explicit form of the ``rung''
$\left[ -i \, G^{\mu \nu}_{rr}(Q) \hat{K}_\mu \hat{K}_\nu \right]$.  
This contraction is gauge invariant, when multiplied by
$\delta (q^0{-}\qpar)$ [as in \Eq{eq:Cdef}],
and we are free to
express $G^{\mu \nu}_{rr}$ in the most convenient gauge, which will be
Coulomb gauge.
With this choice,
\begin {equation}
G^{00}_\ret(Q) = \frac{-1}{\q^2 - \PiL(Q)} \,, \qquad
G^{0i}_\ret(Q) = 0 \,, \qquad
G^{ij}_\ret(Q) = \frac{\delta^{ij} - \hat\q^i \hat\q^j}
	    	       {\q^2 - (q^0)^2 + \PiT(Q)} \,.
\label {eq:3Gs}
\end {equation}
For spacelike $Q \ll T$
[enforced by the delta function in \Eq{eq:Cdef}],
\begin{eqnarray}
\label{eq:Pi}
\PiL(Q)  &=&
	\mD^2 \left[ -1 + \frac{q^0}{2q} \ln \frac{q+q^0}{q-q^0}
	-i \frac{\pi q^0}{2q} \right] ,
\\
\PiT(Q) & = & \mD^2 \, \frac{(q^2-q_0^2)}{2q^2} \left[ 
	\frac{q_0^2}{q^2-q_0^2} + \frac{q^0}{2q} \ln \frac{q+q^0}{q-q^0}
	-i \frac{\pi q^0}{2q} \right] ,
\end{eqnarray}
with $q \equiv |\q|$, and $\mD$ the leading-order Debye mass.
The overall signs in \Eq{eq:3Gs} reflect our metric convention;
in particular, $g^{00} = -1$ appears in the numerator of $G^{00}$.
Recalling that $\hat{K}^\mu$ is a null vector with unit time component,
using \Eq{eq:Grr-def} and \Eq{eq:rho-def}, and approximating 
$n_b(q^0) \simeq T/q^0$ (valid at leading order for $q^0 \sim gT$), 
one finds
\begin{equation}
\label{eq:Grr}
    -i \, G^{\mu \nu}_{rr}(Q) \, \hat{K}_\mu \hat{K}_\nu
    	= \frac{\pi \mD^2 T}{2 q}
	\left\{
	    \frac{2}{|q^2 - \PiL|^2} +
	    \frac{[1-(q^0/q)^2][1-(\qpar/q)^2]}
	{|q^2-(q^0)^2 + \PiT|^2} \right\}
	\, .
\end{equation}
This completes the derivation of \Eq{eq:result2}.

\section{Discussion}
\label{sec:discuss}

\subsection{Symmetry}

In thermal equilibrium, the Wightman current-current correlator
(\ref {eq:Wmunu}) is symmetric under interchange of indices,
$W^{\nu\mu}(K) = W^{\mu\nu}(K)$.
This relation should follow automatically given the
structure of the diagrams which we summed,
but it is not manifest in our result (\ref{eq:jj_final}) for the correlator.
Hence, it is worthwhile to check that this symmetry nevertheless still holds.

The integral equation (\ref{eq:f2}) controls the dependence of
$f^\mu(\p;\k)$ on $\p_\perp$, for fixed values of $\ppar$ and $k$.
To write this equation more abstractly,
consider the vector space of smooth functions of $\p_\perp$
equipped with the natural inner product
\begin{equation}
    \langle g | f \rangle \equiv
    \int \frac{d^2 \p_\perp}{(2\pi)^2} \> g(\p_\perp)^* \, f(\p_\perp) \, .
\end{equation}
One may define linear operators $\widehat{\delta E}$ and $\hat{\cal C}$
in the obvious way so that the integral equation (\ref {eq:f2})
becomes
\begin{equation}
    | {\cal I}^\mu \rangle =
    \Big[ i \widehat{\delta E} + \hat{\cal C} \, \Big] \, | f^\mu \rangle \, ,
\end{equation}
and \Eq{eq:jj_final} turns into an integral over $\ppar$
with an integrand proportional to
$
    \Re \> \langle {\cal I}^\mu | f^\nu \rangle
$,
\begin {equation}
    W^{\mu\nu}_\LPM(K)
	= \int \frac{d\ppar}{2\pi} \>
	\frac{2 n(p^0{+}k^0) \, [1{+}n(p^0)]}{4\ppar \, (k{+}\ppar)} \:
	\Re \> \langle {\cal I}^\mu | f^\nu \rangle \,.
\label {eq:Wxtra}
\end{equation}
[Explicitly,
$\langle {\cal I}^\mu \rangle \equiv 2 P^\mu {+} K^\mu$,
$
    \hat {\cal C} | f \rangle
    \equiv
    \int d^3\q/(2\pi)^3 \> {\cal C}(\q;\k)
    \left[ f(\p_\perp) - f(\p_\perp{+}\q_\perp) \right]
$,
and $\widehat{\delta E}$ is just a multiplication operator by $\delta E$,
all for fixed given values of $\ppar$ and $k$.]

The operators $\widehat{\delta E}$ and $\hat{\cal C}$ are both
real and symmetric, and hence Hermitian;
for the collision operator $\hat{\cal C}$ this is a consequence of the
$\q_\perp \to -\q_\perp$ symmetry (or rotational invariance)
of the kernel (\ref{eq:Grr}).
Moreover, $(\widehat{\delta E})^2$ is positive definite on the 
space of normalizable functions,
which implies that the operator
$\Big[ i \widehat{\delta E} + \hat{\cal C} \Big]$ is invertible.
Consequently, one may write
\begin{equation}
| f^\mu \rangle = \Big[ i \widehat{\delta E} + \hat{\cal C} \, \Big]^{-1}
	| {\cal I}^\mu \rangle \, .
\end{equation}
If $\cal M$ denotes the real part of the inverse,
\begin {equation}
    \hat{\cal M} \equiv
    \half \Big[ i \widehat{\delta E} + \hat{\cal C} \, \Big]^{-1} +
    \half \Big[ -i \widehat{\delta E} + \hat{\cal C} \, \Big]^{-1} \,,
\end {equation}
then
\begin {equation}
    \Re \, \langle {\cal I}^\mu | f^\nu \rangle
    =
    \langle {\cal I}^\mu | \hat{\cal M} | {\cal I}^\nu \rangle \,,
\end {equation}
and
\begin{equation}
    W^{\mu\nu}_\LPM(K)
    = \int {d\ppar \over 2\pi} \:
	\frac{2 n(p^0{+}k^0) \, [1{+}n(p^0)]}{4\ppar \, (k{+}\ppar)} \:
    \langle {\cal I}^\mu | \hat {\cal M} | {\cal I}^\nu \rangle \, .
\end{equation}
Because the operator $\hat{\cal M}$ is, by construction, real and symmetric,
this form shows that our integral equation does yield a current-current
correlation function which is symmetric under interchange of indices.

\subsection{Transversality}
\label{sec:transverse}

The exact current-current correlator is transverse,
$K_\mu W^{\mu \nu}(K) = 0$,
and this implies that the photon emission rate may be computed
either by summing over all external polarization states
with $g^{\mu\nu} W_{\mu\nu}(K)$, or by summing only over the two
transverse physical polarization states with
$\sum_{a=1,2} \epsilon^{\mu}_{(a)} \epsilon^{\nu}_{(a)} W_{\mu \nu}(K)$.

Our result (\ref {eq:jj_final}) is not manifestly transverse, and it is
certainly appropriate to ask whether this property nevertheless holds.
If we had exactly evaluated all of the diagrams we summed
(and included all ``rainbow'' diagrams in the scalar self-energy),
then our result would be exactly transverse.
However, we used various approximations in evaluating these diagrams
which, in the kinematic region of interest, were valid to leading-order
but neglected effects suppressed by additional powers of $g$.
[In particular, this occurred in the steps leading to 
\Eq{eq:f1} and \Eq{eq:Cdef}
where $\qpar$ dependence in $f^\mu(\p{+}\q;\k)$
and $O(\q^2)$ shifts in the pole position of ${\cal F}(P{+}Q;K)$
were neglected.]
Consequently, it is quite possible that our leading-order
approximation (\ref {eq:jj_final}) to the correlator will
not be exactly transverse.
What is essential, however, is that any non-transverse part of $W^{\mu\nu}$
be parametrically small relative to the transverse part.

To examine this issue,
contract both sides of the integral equation
(\ref {eq:f2}) with $K_\mu$ to obtain
\begin{eqnarray}
\label{eq:K.f-eqn}
    (K^2 + 2P_\mu K^\mu)
    =
    (i \delta E) \, K_\mu f^\mu(\p;\k) +
    \int {d^3\q \over (2\pi)^3} \> {\cal C}(\q)
    \left[ K_\mu f^\mu(\p;\k) -  K_\mu f^\mu(\p{+}\q_\perp;\k) \right] \, .
\end{eqnarray}
If one could neglect the collision term altogether,
then the solution would trivially be
\begin {equation}
    K_\mu f^\mu(\p;\k)
    = {K^2 + 2 P_\mu K^\mu \over i \delta E}
    = i \left[E_\p\, \sgn(\ppar) + E_{\p{+}\k} \, \sgn(\ppar{+}k) \right] ,
\label {eq:K.f-trial}
\end {equation}
where the second equality depends on the pinching-pole condition
$2 p^0 + k^0 = E_\p\, \sgn(\ppar) + E_{\p{+}\k} \, \sgn(\ppar{+}k)$
plus the explicit form $E_\p = \sqrt {\p^2 {+} m_\infty^2}$ for the
quasiparticle dispersion relation which neglects $O(g^3 T)$
corrections due to momentum dependence in the self-energy.
Now, 
\begin {equation}
    E_\p\, \sgn(\ppar) + E_{\p{+}\k} \, \sgn(\ppar{+}k)
    = 2\ppar + k + O(\p_\perp^2 / \ppar) \,,
\end {equation}
and the leading $2\ppar{+}k$ piece cancels exactly when
inserted into the collision integral in \Eq{eq:K.f-eqn}.
Consequently, the collision integral is a small perturbation
relative to the other two terms in \Eq{eq:K.f-eqn}.
Iteratively correcting the trial solution (\ref {eq:K.f-trial}),
one easily sees that $K_\mu f^\mu(\p;\k)$ differs from the
trial solution only by terms of order $g^2 T$.
Because the trial solution (\ref {eq:K.f-trial}) is pure imaginary,
this means that the real part of $K_\mu f^\mu(\p;\k)$ is suppressed
relative to the imaginary part by $O(g^2)$.
And since the current-current correlator (\ref {eq:jj_final})
only depends on the real part of $f^\mu(\p;\k)$, this means
that the non-transverse part of (our approximation to)
$W^{\mu\nu}(K)$ is at most $O(g^2 T^3)$.
Note that it was essential to this argument that the collision term
has both a gain and a loss term, of equal size.

In contrast, the transverse part of $f^\mu(\p;\k)$ is of order
$\p_\perp / \delta E \sim 1/g$ [because both $\delta E$
and the collision operator $\hat {\cal C}$ are $O(g^2 T)$],
which leads to an $O(1)$ transverse contribution to $W^{\mu\nu}(K)$.
[The extra power of $g$ comes from the factor of $2P_\perp^\mu = O(gT)$
which appears in the integrand of \Eq{eq:jj_final} when one specializes
to the transverse part.]
Consequently, the longitudinal part of our result for the
current-current correlator is suppressed by $O(g^2)$ relative
to the transverse part.%
\footnote
    {%
    In fact, our result (\ref {eq:jj_final}) for the current-current
    correlator is exactly transverse when the photon momentum
    is precisely on-shell, $K^2 = 0$, if one uses a quasiparticle dispersion
    relation containing just the first two terms in the large
    momentum expansion,
    $E_\p = |\ppar| + \half (\p_\perp^2 {+} m_\infty^2)/|\ppar|$.
    In this case, $K_\mu f^\mu(\p;\k) = 2 P_\mu K^\mu/i\delta E$
    is identically equal to $i(2\ppar {+} k)$, with no sub-leading corrections.
    Hence the collision term in \Eq{eq:K.f-eqn} vanishes identically,
    as does the real part of $K_\mu f^\mu(\p;\k)$.
    }

\subsection{UV and IR behavior}

We next briefly examine the behavior of the integral equation
(\ref {eq:f2}) in the limits of large and small transverse momentum
$\p_\perp$,
and large and small momentum transfer $\q$,
in order to confirm that the photon emission rate is finite
and depends (at leading order) only on $gT$ scale physics.
Since we have just verified transversality of the correlator,
it is sufficient to consider only the transverse components of $f^\mu(\p;\k)$.
For $\p_\perp = O(gT)$ and $\ppar,k = O(T)$, 
recall that both the collision operator $\hat {\cal C}$
and $\delta E$ are $O(g^2 T)$.
Therefore both the real and imaginary parts of $f^\mu_\perp$
are $O(1/g)$.
Now if $\p_\perp \gg gT$, then the $i\delta E$ term in \Eq{eq:f2}
dominates over $\hat {\cal C}$, leading to
\begin{equation}
    f^\mu_\perp(\p_\perp;\k)
    \sim -2i \> \frac{\p_\perp}{\delta E}
    \sim -4i \> {\ppar \, (\ppar{+}k) \over k} \,
	\frac {\p_\perp}{p_\perp^2} \,.
\end{equation}
Hence $\Im\, f^\mu_\perp$ vanishes as $1/|\p_\perp|$ for large $\p_\perp$.
To determine the asymptotic behavior of the real part, we must
expand in $\hat {\cal C} / \delta E$, leading to
\begin{equation}
    \Re \: f^\mu_\perp(\p_\perp;\k)
    \sim
    - \frac{1}{\delta E} \int {d^3\q \over (2\pi)^3} \> {\cal C}(\q) \:
    \Im\left[ f^\mu_\perp ( \p ) - f^\mu_\perp (\p{+}\q_\perp) \right] \, .
\label{eq:re_part}
\end{equation} 
At large $\q$, the collision kernel
${\cal C}(\q) \propto \int_{q^0} G_{rr}^{\mu \nu}(Q) |_{\qpar=q^0} \sim 1/q^5$.
This renders negligible the contribution from 
$|\q_\perp| \gg |\p_\perp|$.
The contribution to $\Re \: f^\mu_\perp$ from $|\q_\perp| \sim |\p_\perp|$
is of order $\p_\perp / (p_\perp^6)$.
To estimate the contribution from $|\q_\perp| \ll |\p_\perp|$ we may make
the approximation that, on angular averaging,
\begin{equation}
\Big[ f^\mu_\perp(\p) -
f^\mu_\perp(\p{+}\q_\perp)\Big]_{{\rm angle \; avg}\, , \,  
	|\q_\perp| \ll |\p_\perp|}
 	\sim -q_\perp^2 \: \nabla_{\p_\perp}^2 \, f^\mu_\perp ( \p ) \, ,
\end{equation}
to conclude that the contribution from $|\q_\perp| \ll |\p_\perp|$ is at
worst logarithmically sensitive to small $q$
(and cut off at $|\q_\perp| \sim gT$),
so the small $\q_\perp$ contribution changes the
$|\q_\perp| \sim |\p_\perp|$ estimate by at most a logarithm. 
Hence, ignoring logs,
\begin{equation}
\label{eq:dE_expand}
    \Re \: f^\mu_\perp (\p_\perp;\k)
    \sim
	\frac{\ppar^2 \, (k{+}\ppar)^2 \, \mD^2 \, g^2 T \: \p_\perp}
	{k^2 \, p_\perp^6} \, .
\end{equation}
This decreases sufficiently rapidly with
$\p_\perp$ to ensure that the $d^2 \p_\perp$ integration in \Eq{eq:jj_final} 
converges at large $\p_\perp$ and is dominantly sensitive to
$\p_\perp = O(gT)$.
One may easily see that there is no problem at small $\p_\perp$,
because both $\delta E$ and the collision term are regular at
$\p_\perp = 0$.
In fact, $f_\perp^\mu$ vanishes linearly in $\p_\perp$ at small
$\p_\perp$.

The large $\q$ behavior of the gauge field correlator,
$G_{rr}(Q) \sim q^{-5}$, ensures that the collision term is
well behaved at large momentum transfer.
But the small momentum transfer behavior requires a bit more care.
First note that, for all $\p_\perp$, the value
of $f^\mu_\perp(\p;\k)$ varies with $\p_\perp$
on a scale of order $gT$ or larger,
and so is slowly varying on smaller scales.
This follows because the collision term smooths $f^\mu$ on this
scale, and $1/\delta E$ is everywhere smooth on this scale.
[Note that $1/\delta E$ is smooth even when $p_\perp \ll gT$ due to the
thermal mass $m_\infty^2 \sim g^2 T^2$ contribution to $\delta E$,
as seen in \Eq{eq:deltaE}.]
Therefore, for small momentum transfer $\q \ll gT$, we may again
approximate the difference $f^\mu_\perp(\p;\k) - f^\mu_\perp(\p{+}\q_\perp;\k)$,
after angular averaging, by a Laplacian,
\begin{equation}
    \Big[ f^\mu_\perp(\p;\k) - f^\mu_\perp(\p{+}\q_\perp;\k)
    \Big]_{\rm angle\; avg.}
    \sim -q^2_\perp \, \nabla_{\p_\perp}^2 \,  f^\mu_\perp(\p;\k) \, .
\label {eq:fdiff}
\end{equation}
Hence, this difference vanishes like $q_\perp^2$ as $\q_\perp \to 0$.
This is what will control the small $q$ behavior.
As discussed in subsection \ref{sec:softer}, 
for $q \ll gT$, $G^{\mu \nu}_{rr}(Q)$ is
sharply peaked in frequency about $q^0 \simeq 0$.
At small $\q^2$, using \Eq{eq:soft}, one has
\begin{equation}
    \int
	\frac{d \qpar \, dq^0}{(2\pi)^2} \> 2\pi \delta(q^0{-}\qpar)
	\left[ -i \, G_{rr}^{\mu \nu}(Q) \right] 
	\simeq
    \frac{\PT^{\mu \nu}(\q_\perp)}{\q_\perp^2} \, .
\end{equation}
Integrating this over $\q_\perp$ would generate a logarithmic IR divergence
(or rather, logarithmic sensitivity to $g^2 T$ scale physics),
exactly as found in the damping rate $\Gamma$.
However, our collision integrand also contains the difference (\ref {eq:fdiff})
so the relevant small $\q_\perp$ behavior is
\begin{equation}
    \int_{\q^2 \ll g^2T^2}
	 \frac{d^2 \q_\perp}{(2\pi)^2} \,
	 \frac{1}{q^2_\perp}
	\left[ q_\perp^2 \nabla_{\p_\perp}^2 \: f^\mu_\perp(\p;\k) \right]
	\sim
	\nabla_{\p_\perp}^2 \: f^\mu_\perp(\p;\k) 
	\int_{\q^2 \ll g^2T^2} q_\perp \> dq_\perp \, .
\end{equation}
This is IR safe by two powers of $q$,
and confirms the irrelevance of momentum transfers which are
much smaller than $O(gT)$.
As expected, it is a cancellation between the gain and loss terms,
which arise from the gluonic cross-rung and self-energy, respectively,
which leads to this good IR behavior.
A treatment which includes only some of the relevant diagrams
would fail to find this cancellation.
For instance, including the self-energy but failing to
sum ladders would miss the gain term, leaving logarithmic
sensitivity to the $g^2 T$ scale.
This explains the conclusions of \cite{Gelis2,Gelis3}.
(Note that such a treatment also fails to produce
a transverse current-current correlator.)

Finally we remark on the small $\ppar$ behavior.
In this regime, the scalar result (\ref{eq:Wxtra}) behaves as 
$
    \int (d\ppar/\ppar^2) \>
    \Re \, \langle \p_\perp \, | \, f_\perp(\p;\k) \rangle
$, 
with one factor of $1/\ppar$ explicit in \Eq{eq:Wxtra} and one coming
from the Bose distribution $n_b(p^0) \sim T/\ppar$ for near on-shell $P$.
This will lead to a small $\ppar$ divergence
(or more properly, linear sensitivity to the inverse thermal mass)
unless $\Re\: f^\mu_\perp(\p;\k)$
vanishes as $\ppar \to 0$.
This is the case, because $\delta E \sim 1/\ppar$
[{\em cf.} \Eq{eq:deltaE}], which becomes large for small $\ppar$.
When $\delta E$ becomes large, our previous analysis leading to
\Eq{eq:dE_expand} shows that
$\Re \: f^\mu_\perp(\p;\k) \propto \ppar^2$, which renders
the $\ppar$ integral in \Eq{eq:Wxtra} well behaved.
The analysis for $k{+}\ppar \ll T$ is exactly the same.
For the case of fermions instead of scalars,
the expression in \Eq{eq:result1} has one
further explicit inverse power of $\ppar$, but the Fermi statistical
function does not have singular small energy behavior,
so the overall small $\ppar$ (or small $\ppar{+}k$) behavior is the same.

\section{Extensions}
\label{sec:extensions}

\subsection{Fermions}
\label{sec:fermions}

It remains to extend our treatment to the physically relevant
case of quarks; that is, charged, nearly massless fermions
instead of charged scalars.
The entire power counting discussion is essentially unchanged.
The only differences arise from the Dirac structure of
the massless fermion propagator which has the form
\begin {equation}
    \frac{1}{\nott{P}+\nott{\Sigma}(P)} \,,
\end {equation}
with a thermal self-energy $\nott\Sigma(P)$ which
does not respect Lorentz invariance
(because the plasma defines a preferred frame).%
\footnote
    {%
    Alternatively, one may say that the self-energy
    does respect Lorentz invariance but depends on
    the plasma 4-velocity $u^\mu$
    as well as the fermion momentum $P^\mu$.
    For a discussion, see \pcite{Weldon}.
    }
At large momentum, $|\p| \gg gT$, 
\begin{equation}
    [\Re \: \Sigma_\mu(P)]^2
    =
    -[\Re \: \Sigma_0(P)]^2 + [\Re \: \vec{\Sigma}(P)]^2 
    = \frac{m_\infty^2}{2|\p|} \, ,
\end{equation}
with $m_\infty^2 = {1\over4} \, g^2 \cf T^2$, the same asymptotic mass
as in the scalar case.
Therefore the mass shell condition $(P+\Re \, \Sigma)^2=0$
is satisfied for a
timelike 4-momentum, $P^2 + m_\infty^2 = 0$,
and the dispersion relation is the same as for a scalar
with vanishing vacuum mass.

Because we are regarding the zero-temperature fermion mass as negligible,
the theory is chirally invariant, and this chiral symmetry is
unbroken in the high temperature plasma.%
\footnote
    {%
    Non-perturbative effects associated with the $U(1)_A$ axial
    anomaly are irrelevant for our perturbative analysis.
    }
This means that we may consider separately the contributions
to photon emission from right and left handed two-component
(Weyl) fermionic fields.
For the right handed component, the $\gamma$ matrices are replaced
by $2\times2$ matrices $\sigma^\mu$ where
$\sigma^0 \equiv {\bf 1}$ and $\sigma^i$ are the Pauli matrices;
for the left handed components, the relevant matrices
are $\bar\sigma^\mu$ which differ just by changing the sign of $\bar\sigma^0$.
It is sufficient to consider the right handed component, as the unpolarized
emission rate from the left handed component is exactly the same.

Rotational invariance guarantees that the spatial self-energy
$\bSigma(P)$ is parallel to $\p$.
This allows us to decompose the fermionic propagator as
\begin{equation}
    \frac{1}{\sigma^\mu (P+\Sigma)_\mu}
    =
	\frac{\half (\sigma^0 + \sigma \cdot \hat{\p})}
	{|\p{+}\bSigma|-(p^0{+}\Sigma^0)}
	-
	\frac{\half (\sigma^0 - \sigma \cdot \hat{\p})}
	{|\p{+}\bSigma|+(p^0{+}\Sigma^0)}
	\, .
\label {eq:fermion-prop}
\end{equation}
The numerators of \Eq{eq:fermion-prop} are helicity projection operators.
It will be convenient to let $u(\p)$ and $v(\p)$ denote
the positive and negative helicity eigenspinors of $\hat \p \cdot \sigma$,
respectively, normalized to $2|\p|$.
Hence,
\begin{equation}
    |\p|\, (\sigma^0 + \sigma \cdot \hat{\p}) = u(\p) u(\p)^\dagger \, ,
\qquad
    |\p|\, (\sigma^0 - \sigma \cdot \hat{\p}) = v(\p) v(\p)^\dagger \, .
\end{equation}

Only one of the two pieces of the propagator (\ref {eq:fermion-prop})
will contribute to the pinching poles in our ladder diagrams,
so at leading order the piece with a non-pinching pole can simply be dropped.  
To be specific, consider bremsstrahlung, so that $\ppar > 0$.
For frequencies near the pinching pole position,
each fermion propagator $G_{\adv}(P)$ may be approximated by 
\begin{equation}
    G_{\adv}(P) \simeq \frac{u(\p) \, u(\p)^\dagger}
	{2 \ppar \, (p^0-E_\p - i \Gamma/2)}  \,,
\end{equation}
for $\ppar > 0$.
This is exactly the same as the scalar propagator except for the
factor $u \,u^\dagger$ in the numerator.
To obtain an expression as similar as possible
to the scalar case, we may associate
the $u$ and $u^\dagger$ spinor with the vertices on
either side of the propagator, instead of the propagator itself.
The gluon vertex is essentially unchanged, because
\begin{equation}
    u(\p)^\dagger \sigma^\mu u(\p{+}\q) \simeq 2 P^\mu \, ,
\end{equation}
with $p^0 = |\p|$, up to $O(|\q|)$ corrections which may be dropped.
For the external photon vertex we need only the transverse components.
It is convenient to work in terms of the two circular polarizations, 
$\sigma^+ = (\sigma^1 {+} i \sigma^2)/\sqrt{2}$ and
$\sigma^- = (\sigma^1 {-} i \sigma^2)/\sqrt{2}$, for which
\begin{equation}
\left| u^\dagger(\p{+}\k) \, \sigma^+ u(\p) \right|^2 = 
	\frac{2 (\ppar{+}k)^2}{\ppar\, (\ppar{+}k)}\, \p_\perp^2 \, , 
\qquad
\left| u^\dagger(\p{+}\k) \, \sigma^- u(\p) \right|^2 = 
	\frac{2 \ppar^2}{\ppar\, (\ppar{+}k)}\, \p_\perp^2 \, .
\end{equation}
(Switching to the other helicity just interchanges the roles of
$\sigma^+$ and $\sigma^-$.)
The corresponding scalar quantity is just $(2\p_\perp)^2$.

In the fermionic case, each diagram has an overall statistical factor of
$
    n_f(k{+}p) n_f(-p) = n_f(k{+}p) \, [1 {-} n_f(p)]
$,
replacing the factors of
$
    n_b(k{+}p) \, [1 {+} n_b(p)]
$
in the scalar case.
Hence,
the leading order contribution of a right (or left) handed fermion to the
photon emission rate, summed over polarizations,
differs from the scalar contribution by a ratio of
\begin{equation}
    \frac{{\rm fermion}}{{\rm scalar}} = 
    \left[ \frac{(\ppar{+}k)^2 + \ppar^2}{2\ppar \, (\ppar{+}k)} \right]
    \left[
	{n_f(k{+}\ppar) \, [1 {-} n_f(\ppar)] 
	\over n_b(k{+}\ppar) \, [1 {+} n_b(\ppar)]}
    \right] .
\end{equation}
A Dirac fermion has two
chiralities so there is a further factor of 2.
Exactly the same result is obtained in the case of \ipa.
(Note that for \ipa, the sign of the first factor is
negative, as $-k<p<0$; but this compensates for a sign
difference in the population functions, because
$[1{+}n_b(p)] = - n_b(-p)$, whereas $[1{-}n_f(p)] = + n_f(-p)$.)
This completes the derivation of the fermionic result in \Eq{eq:result1}.

\subsection {Off-shell photon production}

For some purposes, off-shell photon production can also be of interest.
For instance, the behavior of $W_{\mu \nu}(K)$ for timelike $K$
is important for dilepton production. 
Our analysis up to this point never assumed that the
photon momentum is exactly null;
for $K^2 \ne 0$ one must merely include the appropriate photon frequency
shift $k^0{-}k$ when evaluating the energy shift~(\ref{eq:dE_is}).
This has one noteworthy effect, however.
With a null photon momentum,
the energy shift $\delta E$ is strictly negative for pair
annihilation [$\ppar(\ppar{+}k) < 0$] and strictly positive for
bremsstrahlung [$\ppar(\ppar{+}k)>0$].
For a non-null photon momentum, $k^0{-}k$ must be added to $\delta E$
and this implies that, for some range in $\ppar$,
the energy shift $\delta E$ can vanish for a particular value of $\p_\perp^2$.
This is the point where one or the other of the processes
(bremsstrahlung or pair annihilation) is kinematically allowed
without any accompanying soft scattering off the plasma.

When $g^2 T \ll |k^0{-}k| \ll T$,
this kinematically allowed point occurs when $gT \ll |\p_\perp| \ll T$.
In this regime, our treatment is valid but needlessly complicated.
At leading order, the existence of soft scattering processes is unimportant,
and one may drop the collision term altogether and merely
apply an appropriate $i \epsilon$ prescription to deal with the pole
in $f^\mu(\p_\perp;\ppar,k)$ where $\delta E$ vanishes.
This results in a standard $2 \leftrightarrow 1$ particle
pair annihilation ($k^0 > k$) or DIS ($k^0 < k$) treatment.

If $|k^0 {-} k| \sim T$, then the value where $\delta E$ vanishes occurs,
for typical $\ppar$, at $\p_\perp^2 \sim T^2$.
Near-collinear processes (with or without associated soft scatterings)
cease to dominate the emission rate.
In this case, all our $\p_\perp^2 \ll \ppar^2$ approximations break down,
and our treatment not only can be superseded by a simpler one, it must be.

Finally, if $|k^0{-}k| \lsim O(g^2 T)$, then the detailed structure of the
collision terms remains relevant, and one must solve our full
integral equation (\ref{eq:result2}) to obtain the correct leading
order emission rate.

\subsection {Softer photons}

Up to this point in our analysis, we have focused on
the case of a hard photon momentum, $k \sim T$.  
We now explore the modifications required when this condition is relaxed
and the photon momentum becomes small compared to the temperature.%
\footnote
    {%
    If the photon momentum is instead large compared to the temperature,
    $k \gg T$, then the emission rate becomes exponentially
    small due to the statistical distribution functions,
    but all of our results remain valid.
    }
There are two main changes.  
First, the degree of collinearity needed between the quark and the photon
in order for the pinching-pole approximation to be valid is modified.
Since the thermal width is $O(g^2 T)$, one might expect that
$\delta E \sim g^2 T$ would be required for the poles to pinch.
As $\delta E \sim k \, (\p_\perp^2/\ppar^2)$, smaller $k$ means that 
$\p_\perp^2$ may be larger.
If $k \sim g^2 T$,
this would suggest that the poles still pinch for $\p_\perp^2 \sim T^2$.
Since we relied on $\p^2_\perp \ll \ppar^2$ at many points in our analysis,
it appears that our treatment may break down when $k \lsim g^2 T$.

In fact this estimate is too pessimistic.  As $k$ becomes smaller, the
cancellation we found which limits sensitivity to very-soft
scattering events with $|\q| \ll gT$
begins operating at larger values of $|\q|$.
In fact, there is an $O(\q^2 / \p_\perp^2)$ suppression for scatterings
with $\q^2 \ll \p_\perp^2$.
The pinching pole requirement is that $\delta E$ be comparable to $\Gamma$,
but with $\Gamma$ given by the thermal width due to scatterings
with momentum transfer larger than the characteristic $|\p_\perp|$
(not $gT$).
Hence, the effective size of $\Gamma$ is reduced to
$O(g^4 \, T^3 / \p^2_\perp)$, up to logarithms.
The typical value of $\p_\perp^2$, for which $\Gamma \sim \delta E$, is
therefore $\p_\perp^2 \sim g^2 T^2 \sqrt{T/k}$.
This means that our treatment only breaks down when the photon momentum
reaches the scale $g^4 T$, up to logarithms.
This coincides with the inverse mean free path
on which quarks undergo large angle scatterings.
Extending our analysis below this scale would require major changes.

One other change required for sufficiently small $k$ is the
correct inclusion of the modified photon dispersion
relation which results from photon interactions with the plasma.
Forward scattering off the plasma increases the photon frequency so that
$(k^0{-}k) \sim e^2 T^2 / k$ (or $K^2 \sim e^2 T^2$).
This correction is non-negligible and must be included unless
$(k^0{-}k)$ is much smaller than $\delta E$ for the typical 
range $\p_\perp$ relevant to photon production.
The analysis above shows that
$\delta E \sim g^2 \sqrt {k T}$,
and hence corrections to the photon dispersion relation can be
neglected only when $e^2 T^2 / k \ll g^2 \sqrt{kT}$,
or $k/T \gg (e/g)^{4/3}$.%
\footnote
    {%
    Note however that if we were considering {\em gluon} emission,
    then dispersion corrections due to the medium would {\em always} be
    important, even for $k \sim T$, a point which has been
    ignored in some previous treatments \cite{Baier_QCD},
    although not in others \cite{Zakharov_QCD}.
    }
Our results remain valid for
photon momenta in the range $eT \ll k \lsim (e/g)^{4/3} T$,
provided one solves the
integral equation~(\ref {eq:result2}) with the correct photon
dispersion correction included in $\delta E$.
If $k \lsim eT$, then plasma effects so strongly affect the propagation
of the photon that it no longer travels at nearly the speed of light
(and is more properly viewed as a plasmon ---
a charge density wave in the plasma).
All of our analysis breaks down in this region.

\section* {ACKNOWLEDGMENTS}

We would like to thank Francois Gelis for useful discussions.
This work was supported, in part, by the U.S. Department
of Energy under Grant Nos.~DE-FG03-96ER40956
and DE-FG02-97ER41027.

\end{document}